\documentclass[aps,prd,superscriptaddress,showpacs,nofootinbib]{revtex4-2}

\usepackage{graphicx}
\usepackage{amsfonts,amsmath,amssymb,amsthm,bbding}
\usepackage{xcolor}

\usepackage{enumitem}
\usepackage{array}
\usepackage{multirow}
\usepackage{longtable}

\makeatletter
\newcommand\newtag[2]{#1\def\@currentlabel{#1}\label{#2}}
\newcommand{\gev}{\text{ GeV}}
\newcommand{\mev}{\text{ MeV}}
\makeatother

\def\be{\begin{equation}}
\def\ee{\end{equation}}
\def\ba{\begin{eqnarray}}
\def\ea{\end{eqnarray}}

\begin{document}

\title{Scrutinizing the Mass Matrices in Three-Higgs-Doublet Models with Generalized CP Symmetries}

\author{Duarte D. Correia\footnote{ duarte.d.correia@tecnico.ulisboa.pt}}
\affiliation{CFTP, Departamento de F\'{i}sica, Instituto Superior T\'{e}cnico,
Universidade de Lisboa,\\
Avenida Rovisco Pais 1, 1049 Lisboa, Portugal
}
\author{João P. Silva \footnote{jpsilva@cftp.tecnico.ulisboa.pt}}
\affiliation{CFTP, Departamento de F\'{i}sica, Instituto Superior T\'{e}cnico,
Universidade de Lisboa,\\
Avenida Rovisco Pais 1, 1049 Lisboa, Portugal
}
\author{Jorge C. Romão \footnote{jorge.romao@tecnico.ulisboa.pt}}
\affiliation{CFTP, Departamento de F\'{i}sica, Instituto Superior T\'{e}cnico,
Universidade de Lisboa,\\
Avenida Rovisco Pais 1, 1049 Lisboa, Portugal
}

\date{\today}

\begin{abstract}
We investigate three-Higgs-doublet models with a softly broken generalized
CP (GCP) symmetry,
focusing on the Yukawa sector and its compatibility with quark flavor data.
We first show that the 40 GCP-symmetric models previously
identified are not all physically distinct, by constructing
the basis transformations relating equivalent realizations
and identifying spurious parameters.
We then perform numerical fits to the six quark masses and the four
independent parameters of the Cabibbo--Kobayashi--Maskawa matrix,
further reducing the set of viable models.
In contrast to the GCP-symmetric two-Higgs-doublet case,
we find that GCP-symmetric three-Higgs-doublet models can successfully
reproduce all quark masses and mixings.
In total, 22 inequivalent models are compatible with current experimental data,
and representative benchmark points are presented.
Our results establish softly broken GCP-symmetric
three-Higgs-doublet models as phenomenologically viable extensions
of the Standard Model and provide a framework for further studies
of their flavor phenomenology.
\end{abstract}


\maketitle

\section{\label{sec:intro}Introduction}
Despite its success, the Standard Model (SM) is an incomplete theory
of nature, failing for example to explain
the existence of dark matter,
accommodate the observed baryogenesis,
or predict the masses of neutrinos. 
One natural extension of the SM involves the addition
of scalar doublets, so called multiple Higgs-doublet models (NHDM).
In this work, we focus on models with three Higgs doublets,
where we impose invariance under a generalized CP (GCP) transformation
\be
\Phi_a(x) \rightarrow X_{ab} \Phi_b^\ast(x^0, -\vec{x})\, ,
\label{GCP}
\ee
where $X\in SU(3)$.

Applying the GCP symmetry to the 2HDM scalar potential yields only
three classes of scalar potentials.
Extending the symmetry to the Yukawa sector (and adding a soft breaking term),
one finds that only two models are consistent with no null quark masses,
non block-diagonal quark mixing matrix, 
and a non-zero Jarlskog invariant.
The two models are
the usual CP conserving 2HDM with 18 real Yukawa parameters,
and one single GCP symmetric 2HDM with 12 real
Yukawa parameters \cite{Ferreira:2010bm}.
Thus, with a soft breaking of the CP symmetry,
both models could, in principle, be candidates to accommodate
the precise experimental values for the observables.
Henceforth,
the ``observables'' referred to are the six quark masses and the four independent
parameters of the quark mixing matrix (10 observables in all).
It turns out that such a numerical fit is possible in the former case,
but not in the latter \cite{Ferreira:2010bm}.

Could a genuine GCP model fit all observables?
This inquiry was started in \cite{Bree:2024edl},
by extending GCP symmetries to the 3HDM models.
It was found that the scalar sector will provide four classes of scalar potentials,
and that,
extending the symmetry to the Yukawa sector,
a total of 40 unique models \cite{Bree:2024edl} exist,
consistent with no null quark masses,
non block-diagonal quark mixing matrix, 
and a non-zero Jarlskog invariant.

As our first result, we show that not all
40 models of Ref.~\cite{Bree:2024edl} are physically distinct.
Indeed, we will find specific basis transformations that relate
apparently distinct models.
In some cases, the models are identical, even though they seem
to have a different number of parameters (since some parameters
have no physical consequence; they are spurious).

In a notation to be explained below in Eqs.~\eqref{eq:basis}-\eqref{GCP_fermion},
the only non-trivial 2HDM GCP symmetric model has 
$(\theta,\alpha,\beta,\gamma)=
\left(\frac{\pi}{3},\frac{\pi}{3},\frac{\pi}{3},\frac{\pi}{3}\right)$.
As mentioned, such a model does not provide a good numerical fit
to the 10 observables \cite{Ferreira:2010bm}.
The equivalent
in the 3HDM are seven models $(\theta,\alpha,\beta,\gamma)$ =
\{$\left(\frac{\pi}{3},\frac{\pi}{3},0,\frac{\pi}{3}\right)$, 
   $\left(\frac{\pi}{3},\frac{\pi}{3},\frac{\pi}{3},0\right)$, 
   $\left(\frac{\pi}{3},\frac{\pi}{3},\frac{\pi}{3},\frac{\pi}{3}\right)$,
   $\left(\frac{\pi}{3},0,0,\frac{\pi}{3}\right)$,
   $\left(\frac{\pi}{3},0,\frac{\pi}{3},0\right)$,
   $\left(\frac{\pi}{3},0,\frac{\pi}{3},\frac{\pi}{3}\right)$,
   $\left(\frac{\pi}{3},\frac{\pi}{3},0,0\right)$\}, where each contains
18 real Yukawa parameters. These seven belong to the same class of scalar 
potential, named CPc ($\theta=\pi/3$) in Ref.~\cite{Bree:2024edl}.
As a second result, we will show that it is possible to
fit numerically all observables in such models.
These are not the only possibilities, however,
as there are models with as few as 10 real Yukawa parameters. 
Unfortunately, one cannot fit the experimental values for the 10 observables
with any of those minimal models. This will be our third result. 
One could then wonder if any model with fewer parameters than the CPc models
was able to reproduce the experimental results.
As our fourth result, we will show that this is possible in some cases,
but not in others.

In Section \ref{sec:lag}, we present the Lagrangian of the theory.
In Section \ref{sec:gcp}, we present the GCP transformations, as well as
the models that will be fitted.
In Section \ref{sec:yukawa_matrices} we present the Yukawa textures
for each model, the fitted parameters, and the models' predictions.
We will draw our conclusions in Section \ref{sec:concl}.
Appendix~\ref{app:models} contains a full map of the 40 models in \cite{Bree:2024edl}
and their relations.

\section{\label{sec:lag}The Lagrangian}

\subsection{Scalar Sector}

We start with a three Higgs doublets model (3HDM)
with 3 Higgs-doublets $\Phi_i$, of the same hypercharge $Y=1/2$,
which acquire vacuum expectation values (vevs)
\begin{equation}
\langle \Phi_i \rangle
=
\left(
\begin{array}{c}
\langle \phi^+_i \rangle\\
\langle \phi^0_i \rangle
\end{array}
\right)
=
\left(
\begin{array}{c}
0\\
v_i/\sqrt{2}
\end{array}
\right)\, ,
\label{vev}
\end{equation}
where the index $i$ runs from 1 to 3.

The most general, renormalizable, gauge-invariant 3HDM scalar 
potential can be written, without loss of generality as 
\cite{Botella:1994cs,Branco:1999fs,Davidson:2005cw}
\begin{equation}\label{eq:sclptl}
    V_H = Y_{ij}(\Phi_i^\dagger\Phi_j) +
Z_{ij,kl}(\Phi_i^\dagger\Phi_j)(\Phi_k^\dagger\Phi_l)\,\text{,}
\end{equation}
where $Y$ and $Z$ are 3-dimensional rank-2 and rank-4, respectively, tensors in Higgs-doublet space, and $Z_{ij,kl}\equiv Z_{kl,ij}$.
$V_H$ is real, which implies that
\begin{eqnarray}
Y_{ij} &=& Y_{ji}^\ast\, ,
\nonumber\\
Z_{ij,kl}&=& Z_{ji,lk}^\ast  \equiv Z_{lk,ji}^\ast \, .
\label{Hermiticity_coefficients}
\end{eqnarray}
This reduces the amount of independent entries in the tensors to 3 real and 3 complex in $Y$, for a total of 9 real parameters, and 9 real and 18 complex in $Z$, for a total of 45 real parameters.

\subsection{Yukawa Sector}

In the Yukawa sector, the scalar-quark interactions can be written as 
\be
- {\cal L}_Y =
\bar{q}_L
(\Gamma_1 \Phi_1 + \Gamma_2 \Phi_2 + \Gamma_3 \Phi_3)
n_R
+
\bar{q}_L
(\Delta_1 \tilde{\Phi}_1 + \Delta_2 \tilde{\Phi}_2 + \Delta_3 \tilde{\Phi}_3)
p_R
+ \textrm{H.c.},
\label{eq:yuk}
\ee
where $q_L = (p_L, n_L)^\top$ ($n_R$ and $p_R$) is a vector in the 3-dimensional
generation space of left-handed doublets
(right-handed charge $-1/3$ and $+2/3$) quarks,
and $\Tilde{\Phi}_k = i \sigma_2 {\Phi}^*_k$, with $\sigma_2$ the second Pauli matrix.  
$\Gamma_k$ and $\Delta_k$ ($k=1,2,3$) are completely
general $3 \times 3$ complex matrices,
and H.c. stands for Hermitian conjugate.

After spontaneous symmetry breaking (SSB),
the Lagrangian acquires mass terms 
\begin{equation}
        -\mathcal{L}_\textrm{mass} = \overline{n}_L\, M_d\, n_R + \overline{p}_L\, M_u\, p_R + \text{H.c., }
\end{equation}
where we have
\begin{equation}
M_d \equiv \sum_i v_i \Gamma_i/\sqrt{2}\, ,
\ \ \ \
M_u \equiv
\sum_i v_i^*\Delta_i/\sqrt{2}\, ,
\end{equation}
and $\sum_i |v_i|^2 \equiv v^2 = (\sqrt{2} G_F)^{-1}$ ($i=1,2,3$).

\subsection{CKM Matrix and CP Violation}

Since the fields we are working with are not mass eigenstates, the mass matrices 
$M_d$ and $M_u$ will not be diagonal.
But we can always bi-diagonalize these matrices by performing a unitary
change of basis in the quark fields, which leaves the kinetic terms in the
Lagrangian unchanged
\begin{equation}
    \begin{split}
        \overline{n}_L = \overline{d}_L U_{d_L}^\dagger\, , \ \ \ \ \ 
\overline{p}_L = \overline{u}_L U_{u_L}^\dagger\,\text{, }\\
        n_R = U_{d_R} d_R\, , \ \ \ \ \ 
p_R = U_{u_R} u_R\,\text{. }
    \end{split}
\end{equation}
The mass matrices then become
\begin{equation}
    \begin{split}
        U_{d_L}^\dagger M_d U_{d_R} &=D_d \equiv \text{diag}(m_d, m_s, m_b)\,\text{, }\\
        U_{u_L}^\dagger M_u U_{u_R} &=D_u \equiv \text{diag}(m_u, m_c, m_t)\,\text{.}
    \end{split}
\end{equation}

These changes of basis have some consequences elsewhere in the Lagrangian.
If we take a look at the interaction between the physical fields and the $W^+$ boson, 
\begin{equation}\label{eq:22}
    \mathcal{L}_{W} \supset \frac{i g}{\sqrt{2}}\, 
\overline{u}_L (U_{u_L}^\dagger U_{d_L}) \gamma^\mu d_L\, W_\mu^+ \, ,
\end{equation}
one notices the appearance of a $3\times3$ unitary matrix,
the Cabibbo-Kobayashi-Maskawa (CKM) matrix \cite{Cabibbo:1963yz,Kobayashi:1973fv},
$V\equiv U_{u_L}^\dagger U_{d_L}$.
It is indicative that the physical basis is distinct from the interaction basis,
and describes quark mixing, which is responsible for all CP-violating 
phenomena in the SM.

We can also define the Hermitian matrices
\begin{equation}\label{eq:HH}
    \begin{split}
        H_d \equiv  M_d  M_d^\dagger
=U_{d_L} D_d^2 U_{d_L}^\dagger \, ,\\
        H_u \equiv  M_u  M_u^\dagger
=  U_{u_L} D_u^2 U_{u_L}^\dagger\, ,
    \end{split}
\end{equation}
showing that the left-handed transformations are the
matrices that diagonalize $H_d$ and $H_u$. 

There is one simple constraint, regarding CP violation,
that these matrices must respect.
Experiments in $B$ decays prompt the conclusion that there must
be CP violation in the CKM matrix, which is related to 
the Jarlskog invariant $J_{CP}$
\cite{Jarlskog:1985ht,Jarlskog:1985cw}.
In terms of the CKM matrix elements this invariant is
given by \cite{ParticleDataGroup:2024cfk}
\begin{equation}
    J_{CP} = \text{Im}(V_{us}V_{cb}V^*_{ub}V^*_{cs})
= (3.12^{+0.13}_{-0.12})\times 10^{-5}\,\text{.}
\end{equation}
One can construct another related CP-violating, basis invariant, quantity
\cite{Bernabeu:1986fc,Jarlskog:1985ht,Jarlskog:1985cw}
\begin{equation}
    \begin{split}
        J = \text{Tr}\left[ H_u, H_d\right]^3 
        &= 6i(m_t^2-m_c^2)(m_t^2-m_u^2)(m_c^2-m_u^2)\\
        &\times (m_b^2-m_s^2)(m_b^2-m_d^2)(m_s^2-m_d^2) J_{CP}\,\text{.}
    \end{split}
\end{equation}
Note that $J=0$ if and only
if the CKM matrix conserves CP, i.e. $J_{CP}=0$, or if there is mass degeneracy
in either the up or down quark sectors, both of which are ruled out experimentally.
Therefore, whatever textures we find for the $\Gamma_k$ and $\Delta_k$ matrices,
$\text{Tr}\left[ H_u, H_d\right]^3$ must not vanish.

\section{\label{sec:gcp}GCP Symmetry}

Under a GCP transformation, the scalar fields transform as 
\be
\Phi_a\rightarrow X_{ab} \Phi_b^\ast,
\ee
where $X$ is a $3\times3$ unitary matrix acting in scalar doublet space, and
we simplified the notation from Eq.~\eqref{GCP}. 
We could redefine our fields by
\be
\Phi\rightarrow W \Phi',
\ee
where, in the new basis, the GCP transformation matrix is given by
\be
X' = W^\dagger X W^\ast \Leftrightarrow X' \cong X ,
\ee
with $W$ a unitary matrix. Note that this is not a change of basis, but a {\it congruence} transformation.
Since unitary transformations do not change the physics of the theory,
and in order to help solve the invariance conditions that arise, 
we may use the fact (proven by Ecker, Grimus, and Neufeld \cite{Ecker:1987qp}) that there always exists a unitary matrix
$W$ such that the matrix $X$
may be brought to the form
\begin{equation}\label{eq:basis}
    W^\top X\,W =X_\theta = \begin{bmatrix}
        \cos{\theta} & \sin{\theta} & 0 \\
        -\sin{\theta} & \cos{\theta} & 0 \\
        0 & 0 & 1
    \end{bmatrix} \equiv R_\theta\oplus 1\, \text{,}
\end{equation}
where the $\oplus$ symbol stands for direct sum, and $0\leq\theta\leq\pi/2$.

For the quark fields, the GCP transformations take the form
\ba
q_L &\rightarrow X_\alpha  \gamma^0 C q^\ast_L,\nonumber \\
n_R &\rightarrow X_\beta  \gamma^0 C n^\ast_R,\nonumber \\
p_R &\rightarrow X_\gamma  \gamma^0 C p^\ast_R,
\label{GCP_fermion}
\ea
where $\gamma^0$ ($C$) is the Dirac (charge-conjugation) matrix,
and $X_\alpha$, $X_\beta$ , and $X_\gamma$ belong to $SU(3)$.
With a suitable basis choice \cite{Ecker:1987qp},
these can be changed into the simplified form of Eq.~\eqref{eq:basis}, where,
as before, $0 \leq \alpha,\beta,\gamma \leq \pi/2$.

The scalar potential in Eq.~\eqref{eq:sclptl} is invariant
under the GCP transformation in Eq.~\eqref{GCP} if and only if
\begin{equation}\label{eq:sclinv}
    \begin{split}
    & Y_{ab}^* = (X_\theta)_{\alpha a}^* Y_{\alpha\beta} (X_\theta)_{\beta b}\, , \\
    & Z_{ab,cd}^* = (X_\theta)_{\alpha a}^* (X_\theta)_{\gamma c}^*
Z_{\alpha\beta,\gamma\delta} (X_\theta)_{\beta b} (X_\theta)_{\delta d}\,\text{. }
    \end{split}
\end{equation}
Solving Eqs.~\eqref{eq:sclinv} in the basis of Eq.~\eqref{eq:basis},
there are just four unique scalar potentials, denoted by CPa ($\theta=0$),
CPb ($\theta=\pi/2$), CPc ($\theta=\pi/3$), and CPd ($\theta \neq 0, \pi/2, \pi/3$) \cite{Bree:2024edl}.

Turning our attention to the Yukawa sector,
the Yukawa Lagrangian in Eq.~\eqref{eq:yuk}
is invariant under the GCP transformations
in Eqs.~\eqref{GCP_fermion} if and only if
\begin{equation}\label{eq:qrkinv}
    \begin{split}
        \Gamma_b^* &= X_\alpha^\dagger
(X_\theta)_{ab} (\Gamma_a) X_\beta\, , \\
        \Delta_b^* &=  X_\alpha^\dagger
(X_\theta)_{ab}^*(\Delta_a)  X_\gamma \,\text{.}
    \end{split}
\end{equation}
Since we are taking all $X$ matrices to be real,
the condition for the charged $+2/3$ quark matrices $\Delta$ will
yield the same equations as the charged $-1/3$ quark matrices $\Gamma$,
under the substitution $\beta\xrightarrow{}\gamma$.

Computing all combinations of angles $\theta$, $\alpha$, $\beta$, and $\gamma$,
and keeping only those that are consistent with no null quark masses,
non block-diagonal quark mixing matrix, 
and a non-zero Jarlskog invariant,
Ref.~\cite{Bree:2024edl} found 40 unique textures for the
Yukawa matrices \cite{Bree:2024edl}.
For completeness,
we include them in Table~\ref{tab:fixedmodel}
and Table~\ref{tab:rangemodel}.
For each model we include the allowed values for $\theta$,
along with the amount of real independent parameters in
the down-type and up-type Yukawa matrices,
and whether or not the model was found to fit the data.
An asterisk means that, for that Yukawa matrix,
the third quark decouples from the first two and a diamond
means that the quarks couple only to $\Phi_3$. 

\begin{table}[htp!]
    \centering
    \begin{tabular}{c|c|c|c|c}
     \hline
\hline
\multirow{2}{*}{$(\theta,\alpha,\beta,\gamma)$}
     & \multirow{2}{*}{Range for $\theta$} 
     & \# of real param. & \# of real param. & Experimental\\
     &  & in down-type Yukawa & in up-type Yukawa & Data Fit\\
\hline
\\[-0.9em]
$(0,0,0,0)$ & 0 & 27 & 27 & \Checkmark
\\[0.2em]
\hline
\\[-0.9em]
$(\frac{\pi}{2},0,0,\frac{\pi}{2})$ & \multirow{7}{*}{$\frac{\pi}{2}$} & $9^\diamond$ & 15 
& \Checkmark\\[0.2em]
$(\frac{\pi}{2},0,\frac{\pi}{2},0)$ &  & 15 & $9^\diamond$
& \Checkmark\\[0.2em]
$(\frac{\pi}{2},0,\frac{\pi}{2},\frac{\pi}{2})$ &  & 15 & 15
& \Checkmark\\[0.2em]
$(\frac{\pi}{2},\frac{\pi}{2},0,0)$ &  & 15 & 15
& \Checkmark\\[0.2em]
$(\frac{\pi}{2},\frac{\pi}{2},0,\frac{\pi}{2})$ &  & 15 & 13
& \Checkmark\\[0.2em]
$(\frac{\pi}{2},\frac{\pi}{2},\frac{\pi}{2},0)$ &   & 13& 15
& \Checkmark\\[0.2em]
$(\frac{\pi}{2},\frac{\pi}{2},\frac{\pi}{2},\frac{\pi}{2})$ &   & 13& 13
& \Checkmark\\[0.2em]
\hline
\\[-0.9em]
$(\frac{\pi}{3},\frac{\pi}{3},0,\frac{\pi}{3})$ &  \multirow{3}{*}{$\frac{\pi}{3}$} & 9 & 9 & \Checkmark \\[0.2em]
$(\frac{\pi}{3},\frac{\pi}{3},\frac{\pi}{3},0)$ &   & 9 & 9 & \Checkmark \\[0.2em]
$(\frac{\pi}{3},\frac{\pi}{3},\frac{\pi}{3},\frac{\pi}{3})$ &   & 9 & 9
& \Checkmark \\[0.2em]
\hline
\\[-0.9em]
$(\frac{\pi}{4},\frac{\pi}{4},0,\frac{\pi}{2})$ &  \multirow{6}{*}{$\frac{\pi}{4}$} & 9 & 7
& \Checkmark\\[0.2em]
$(\frac{\pi}{4},\frac{\pi}{4},\frac{\pi}{2},0)$ &   & 7 & 9
& \Checkmark\\[0.2em]
$(\frac{\pi}{4},\frac{\pi}{4},\frac{\pi}{4},\frac{\pi}{2})$ &  & 7 & 7
& \Checkmark\\[0.2em]
$(\frac{\pi}{4},\frac{\pi}{4},\frac{\pi}{2},\frac{\pi}{4})$ &   & 7 & 7
& \Checkmark\\[0.2em]
$(\frac{\pi}{4},\frac{\pi}{4},\frac{\pi}{2},\frac{\pi}{2})$ &   & 7 & 7
& \XSolid\\[0.2em]
$(\frac{\pi}{4},\frac{\pi}{2},\frac{\pi}{4},\frac{\pi}{4})$ &   & 7 & 7
& \XSolid\\[0.2em]
\hline
\\[-0.9em]
$(\frac{\pi}{5},\frac{2\pi}{5},\frac{\pi}{5},\frac{2\pi}{5})$ &  \multirow{2}{*}{$\frac{\pi}{5}$} & 5 & $5^*$ & \XSolid \\[0.2em]
$(\frac{\pi}{5},\frac{2\pi}{5},\frac{2\pi}{5},\frac{\pi}{5})$ & & $5^*$ & 5
& \XSolid \\[0.2em]
\hline
\\[-0.9em]
$(\frac{2\pi}{5},\frac{\pi}{5},\frac{2\pi}{5},\frac{\pi}{5})$ &  \multirow{2}{*}{$\frac{2\pi}{5}$} & 5 & $5^*$ & \XSolid\\[0.2em]
$(\frac{2\pi}{5},\frac{\pi}{5},\frac{\pi}{5},\frac{2\pi}{5})$ & & $5^*$ & 5
& \XSolid\\[0.2em]
\hline
\hline
\end{tabular}
\caption{\label{tab:fixedmodel}All 21 GCP-symmetric 3HDM compatible
with non zero and non-degenerate quark masses as well as a
non-vanishing Jarlskog invariant where all transformation angles are fixed.
Taken from Ref.~\cite{Bree:2024edl}. See text for further details.}
\end{table}

\begin{table}[htp!]
    \centering
    \begin{tabular}{c|c|c|c|c}
     \hline
\hline
\multirow{2}{*}{$(\theta,\alpha,\beta,\gamma)$}
     & \multirow{2}{*}{Range for $\theta$} 
     & \# of real param. & \# of real param. & Experimental \\
     &  & in down-type Yukawa & in up-type Yukawa & Data Fit \\
\hline
\\[-0.9em]
$(\theta,0,0,\theta)$ & \multirow{4}{*}{$\left(0, \frac{\pi}{2}\right)$}  & $9^\diamond$ & $9$ & \Checkmark\\[0.2em]
$(\theta,0,\theta,0)$ &   & $9$& $9^\diamond$ & \Checkmark\\[0.2em]
$(\theta,0,\theta,\theta)$ &   & $9$& $9$ & \Checkmark\\[0.2em]
$(\theta,\theta,0,0)$ &   & $9$& $9$
& \Checkmark\\[0.2em]
\hline
\\[-0.9em]
$(\theta,\theta,0,\theta)$ & \multirow{3}{*}{$\left(0, \frac{\pi}{2}\right)\setminus\{\frac{\pi}{3}\}$}   & $9$ & $7$ & \Checkmark\\[0.2em]
$(\theta,\theta,\theta,0)$ &  & $7$ & $9$ & \Checkmark\\[0.2em]
$(\theta,\theta,\theta,\theta)$ &   & $7$& $7$
& \Checkmark\\[0.2em]
\hline
\\[-0.9em]
$(\theta,\theta,0,2\theta)$ &  \multirow{6}{*}{ $\left(0, \frac{\pi}{4}\right)$} & $9$ & $5$ & \Checkmark\\[0.2em]
$(\theta,\theta,2\theta,0)$ &  & $5$ & $9$ & \Checkmark\\[0.2em]
$(\theta,\theta,\theta,2\theta)$ & & $7$ & $5$ & \Checkmark\\[0.2em]
$(\theta,\theta,2\theta,\theta)$ &  & $5$ & $7$ & \Checkmark\\[0.2em]
$(\theta,\theta,2\theta,2\theta)$ &  & $5$ & $5$ & \XSolid\\[0.2em]
$(\theta,2\theta,\theta,\theta)$ &  & $5$ & $5$
& \XSolid\\[0.2em]
\hline
\\[-0.9em]
$(\theta,\theta,0,\pi-2\theta)$ &  \multirow{6}{*}{ $\left(\frac{\pi}{4}, \frac{\pi}{2}\right)\setminus\{\frac{\pi}{3}\}$}  & $9$ & $5$ & \Checkmark\\[0.2em]
$(\theta,\theta,\pi-2\theta,0)$ &  & $5$ & $9$ & \Checkmark\\[0.2em]
$(\theta,\theta,\theta,\pi-2\theta)$ & & $7$ & $5$ & \Checkmark\\[0.2em]
$(\theta,\theta,\pi-2\theta,\theta)$ &  & $5$ & $7$ & \Checkmark\\[0.2em]
$(\theta,\theta,\pi-2\theta,\pi-2\theta)$ &  & $5$ & $5$ & \XSolid\\[0.2em]
$(\theta,\pi-2\theta,\theta,\theta)$ &  & $5$ & $5$ & \XSolid\\[0.2em]
\hline
\hline
\end{tabular}
\caption{\label{tab:rangemodel}All 19 GCP-symmetric 3HDM
compatible with non zero and non-degenerate quark masses as
well as a non-vanishing Jarlskog invariant where at least
one of the transformation angles can vary in a range.
Taken from Ref.~\cite{Bree:2024edl}.}
\end{table}

In Table~\ref{tab:rangemodel}, the allowed values for $\theta$ 
were chosen such that $\theta$, $\alpha$, $\beta$, and $\gamma$
lie simultaneously between 0 and $\pi/2$, as well as not 
allowing any combination already listed in Table~\ref{tab:fixedmodel}.
In the following section, we present the fitting procedure, and 
a numerical fit for some of the models,
namely every model invariant under CPc, 
as well as most models invariant under CPd.
We also discuss how the similarities between Yukawa textures 
across different models allow us to extrapolate whether a model
is able to provide a good fit or not.

\section{\label{sec:yukawa_matrices}Yukawa Textures}

\subsection{`Boundary' Models}

We split the models, labelled by the transformation angles 
$(\theta,\alpha,\beta,\gamma)$, into Tables~\ref{tab:fixedmodel} and~\ref{tab:rangemodel}
according to whether or not
the model requires specific values for every angle, 
respectively.

Consider the following three models, 
$(\pi/3,\pi/3,\pi/3,0)$,
$(\theta,\theta,\theta,0)$,
and $(\theta,\theta,\pi-2\theta,0)$, where in the last two,
$\theta$ can be any value in the respective interval,
stated in Table~\ref{tab:rangemodel}.
For these three models, the up-type textures are identical, 
but the down-type ones are different, though similar. 
The down-type textures are, respectively, 
\ba
&&
\Gamma_1 =
\left(
\begin{array}{ccc}
 i a_{11} & i a_{12} & a_{13} \\
 i a_{12} & -i a_{11} & a_{23} \\
 a_{31} & a_{32} & 0
\end{array}
\right),
\ \ \
\Gamma_2 =
\left(
\begin{array}{ccc}
 i a_{12} & -i a_{11} & -a_{23} \\
 -i a_{11} & -i a_{12} & a_{13} \\
 -a_{32} & a_{31} & 0
\end{array}
\right),
\ \ \
\Gamma_3 =
\left(
\begin{array}{ccc}
c_{11} & c_{12} & 0 \\
-c_{12} & c_{11} & 0 \\
0 & 0 & c_{33}
\end{array}
\right),
\nonumber \\[1em]
&&
\Gamma_1 =
\left(
\begin{array}{ccc}
 0 & 0 & a_{13} \\
 0 & 0 & a_{23} \\
 a_{31} & a_{32} & 0
\end{array}
\right),
\ \ \
\Gamma_2 =
\left(
\begin{array}{ccc}
 0 & 0 & -a_{23} \\
 0 & 0 & a_{13} \\
 -a_{32} & a_{31} & 0
\end{array}
\right),
\ \ \
\Gamma_3 =
\left(
\begin{array}{ccc}
c_{11} & c_{12} & 0 \\
-c_{12} & c_{11} & 0 \\
0 & 0 & c_{33}
\end{array}
\right),
\nonumber \\[1em]
&&
\Gamma_1 =
\left(
\begin{array}{ccc}
 i a_{11} & i a_{12} & a_{13} \\
 i a_{12} & -i a_{11} & a_{23} \\
 0 & 0 & 0
\end{array}
\right),
\ \ \
\Gamma_2 =
\left(
\begin{array}{ccc}
 i a_{12} & -i a_{11} & -a_{23} \\
 -i a_{11} & -i a_{12} & a_{13} \\
 0 & 0 & 0
\end{array}
\right),
\ \ \
\Gamma_3 =
\left(
\begin{array}{ccc}
0 & 0 & 0 \\
0 & 0 & 0 \\
0 & 0 & c_{33}
\end{array}
\right).
\nonumber
\ea

We notice that the second set of textures can be obtained 
from the first by setting 
$a_{11}$ and $a_{12}$ to zero,
whereas we can similarly obtain the third set
by setting $a_{31}$, $a_{32}$, $c_{11}$, and $c_{12}$
to zero. It is not a coincidence that $\pi/3$ is a boundary
point for the allowed range of $\theta$ values in the
$(\theta,\theta,\theta,0)$
and $(\theta,\theta,\pi-2\theta,0)$ models.
In fact, nothing stops us from setting $\theta=\pi/3$ in these models,
but then the respective textures will not be the
most general matrices
which are invariant under Eqs.~\eqref{GCP} and \eqref{GCP_fermion} with $X_\theta=X_\alpha=X_\beta=X_{\pi/3}$,
and $X_\gamma=X_0\equiv{\bold 1}$, hence why the boundary
values for $\theta$ are treated separately.

We can then define the concept of {\it boundary models}.
For a given model, the Yukawa textures will be entirely
determined by a set of parameters $\{x_{ij}\}$. A boundary model is
defined as a model where we can set extra conditions on 
a subset of the parameters,
$\{y_{ij}\}$, e.g.~setting them to zero or equal to each other,
such that the remaining texture is identical 
to a different model's texture.

We can think of each model as a hypersurface in parameter space,
and by setting some combination of the parameters to zero,
effectively slicing the surface, it is possible
that the remaining subspace perfectly coincides with the hypersurface
defined by a different model, with fewer parameters.
It can be shown that this happens if and only if 
the larger model is
listed in Table~\ref{tab:fixedmodel}, the smaller in
Table~\ref{tab:rangemodel}, and the transformation angles
can become identical for some value of $\theta$, i.e.~the values at the boundaries,
hence the name boundary model. 
For example, the $(0,0,0,0)$, 
$(\pi/2,\pi/2,\pi/2,\pi/2)$, and $(\pi/3,\pi/3,\pi/3,\pi/3)$ models are boundary models for the $(\theta,\theta,\theta,\theta)$ model, and thus the 
latter's parameter space is fully contained in each of the previous three.

This is a helpful fact because if a model with a lower number of parameters
fits the data well, then its boundary models, however many there are,
will necessarily be able to fit the data at least as well.
This is not an equivalence, however, since
a model that fits the data poorly could in principle have a boundary model
that is able to fit the data given the new degrees of freedom in the 
extra parameters.

\subsection{Removing Redundancies}

Consider the set of models where at least one angle equals
$\pi-2\theta$, listed in Table~\ref{tab:rangemodel}.
In principle, we could have picked any value for $\theta$,
but the specific range was chosen such that every transformation angle
lies between 0 and $\pi/2$. 
Take for example $(\theta,\theta,\pi-2\theta,0)$, and let $\theta=\pi/5$, the transformation angles then would become
$(\pi/5,\pi/5,3\pi/5,0)$, $\beta$ sits outside the allowed range.
But, as it was proven by Ecker,
Grimus, and Neufeld \cite{Ecker:1987qp}, there is a basis
$n_R\rightarrow W n'_R$ where $W^\top X_\beta W = X_{\beta'}$
with $0\leq\beta'\leq\pi/2$.

For any $n\times n$ unitary matrix $X$, the angles of this "standard" form 
are given by the Hermitian matrix
\be
\frac{1}{4}(X+X^\top)^\dagger(X+X^\top)\,,
\ee
which, in our $n=3$ case, has twice degenerate eigenvalue
$\cos^2\theta$, and an eigenvalue 1 \cite{Ecker:1987qp}. 
Since $\cos^2(\pi-2\theta)=\cos^2(2\theta)$, then 
$X_{\pi-2\theta}\cong X_{2\theta}$. This means that,
in principle, any model with at least one angle being 
$\pi-2\theta$ is physically equivalent to the same model under
$\pi-2\theta\rightarrow 2\theta$.

Let
\be
W = 
\begin{pmatrix}
 0 & i & 0 \\
 i & 0 & 0 \\
 0 & 0 & 1
\end{pmatrix}\,,
\label{eq:cong}
\ee
then we have
\be
W^\top\, X_{\pi-2\theta}\, W = X_{2\theta}\,,
\label{eq:cong2}
\ee
which proves they are indeed congruent.

Since unitary transformations don't change the physical
predictions of a theory, we say that two models are physically
equivalent if under a weak-basis transformation (WBT)
\ba
q_L &\rightarrow& W_L\, q'_L\,,\nonumber \\
n_R &\rightarrow& W_{dR}\, n'_R\,,\nonumber \\
p_R &\rightarrow& W_{uR}\, p'_R\,,
\ea
the Yukawa textures become identical. Under a WBT, the
Yukawa matrices become $(k=1,2,3)$
\ba
\Gamma'_k &=& W_L^\dagger\, \Gamma_k\, W_{dR}\,,\nonumber \\
\Delta'_k &=& W_L^\dagger\, \Delta_k\, W_{uR}\,.
\label{eq:WBT}
\ea

Consider the models $(\theta,\theta,\pi-2\theta,0)$ and 
$(\theta,\theta,2\theta,0)$ (or alternatively, $(\theta,\theta,\pi-2\theta,\theta)$ and 
$(\theta,\theta,2\theta,\theta)$).
The up-type textures are identical (in each pair). The down-type textures
are, respectively,
\ba
&&
\Gamma_1 =
\left(
\begin{array}{ccc}
 i a_{11} & i a_{12} & a_{13} \\
 i a_{12} & -i a_{11} & a_{23} \\
 0 & 0 & 0
\end{array}
\right),
\ \ \
\Gamma_2 =
\left(
\begin{array}{ccc}
 i a_{12} & -i a_{11} & -a_{23} \\
 -i a_{11} & -i a_{12} & a_{13} \\
 0 & 0 & 0
\end{array}
\right),
\ \ \
\Gamma_3 =
\left(
\begin{array}{ccc}
0 & 0 & 0 \\
0 & 0 & 0 \\
0 & 0 & c_{33}
\end{array}
\right),
\nonumber \\[1em]
&&
\Gamma_1 =
\left(
\begin{array}{ccc}
 a_{11}  & a_{12} & a_{13} \\
 -a_{12} & a_{11} & a_{23} \\
 0 & 0 & 0
\end{array}
\right),
\ \ \
\Gamma_2 =
\left(
\begin{array}{ccc}
 -a_{12} & a_{11} &-a_{23} \\
 -a_{11} &-a_{12} & a_{13} \\
 0 & 0 & 0
\end{array}
\right),
\ \ \
\Gamma_3 =
\left(
\begin{array}{ccc}
0 & 0 & 0 \\
0 & 0 & 0 \\
0 & 0 & c_{33}
\end{array}
\right).
\nonumber
\ea
Let $W_L=W_{uR}={\bold 1}$ and $W_{dR}=W$ defined in Eq.~\eqref{eq:cong}. Applying Eqs.~\eqref{eq:WBT}, as well as
the substitution $a_{11}\leftrightarrow-a_{12}$, to the Yukawa matrices of $(\theta,\theta,\pi-2\theta,0)$
$\left((\theta,\theta,\pi-2\theta,\theta)\right)$, 
we exactly obtain the Yukawa matrices of 
$(\theta,\theta,2\theta,0)$
$\left((\theta,\theta,2\theta,\theta)\right)$.

These models share a boundary model, $(\pi/4,\pi/4,\pi/2,0)$
$\left((\pi/4,\pi/4,\pi/2,\pi/4)\right)$, which requires
special attention. The up-type textures remain the same,
but the down-type ones are
\ba
&&
\Gamma_1 =
\left(
\begin{array}{ccc}
  a_{11} e^{i\rho} & a_{12} e^{i\sigma} & a_{13} \\
 -a_{12}e^{-i\sigma} & a_{11}e^{-i\rho} & a_{23} \\
 0 & 0 & 0
\end{array}
\right),
\ \ \
\Gamma_2 =
\left(
\begin{array}{ccc}
 -a_{12}e^{-i\sigma} & a_{11}e^{-i\rho} &-a_{23} \\
 -a_{11}e^{i\rho} &-a_{12}e^{i\sigma} & a_{13} \\
 0 & 0 & 0
\end{array}
\right),
\ \ \
\Gamma_3 =
\left(
\begin{array}{ccc}
0 & 0 & 0 \\
0 & 0 & 0 \\
0 & 0 & c_{33}
\end{array}
\right).
\nonumber
\ea

We notice that we can obtain either the $\pi-2\theta$ or the
$2\theta$ textures simply by setting $\rho=\sigma=\pi/2$ or
$\rho=\sigma=0$, respectively. Our argument above proved 
both these cases are physically equivalent, but the same argument would map this texture onto itself. 
The question remains of whether a WBT can arbitrarily set $\rho=\sigma=0$, if so then $(\pi/4,\pi/4,\pi/2,0)$
$\left((\pi/4,\pi/4,\pi/2,\pi/4)\right)$ would be
physically equivalent to $(\theta,\theta,2\theta,0)$
$\left((\theta,\theta,2\theta,\theta)\right)$ as well.
Let $W_L=W_{uR}={\bold 1}$, since the up-type textures are
already equal, and
\be
W_{dR} = 
\begin{pmatrix}
 \cos \eta\ e^{i\zeta} & \sin \eta\ e^{-i\zeta} & 0 \\
 -\sin \eta\ e^{i\zeta} & \cos \eta\ e^{-i\zeta} & 0 \\
 0 & 0 & 1
\end{pmatrix}\,,
\ee
where $\zeta$ satisfies $\cos 2\zeta = 
\frac{1}{a_{11}^2+a_{12}^2}(a_{11}^2 \cos 2\rho + a_{12}^2 \cos 2\sigma)$, 
and $\eta$ satisfies $\tan \eta = 
\frac{a_{11} \sin(\zeta + \rho)}{a_{12} \sin(\zeta + \sigma)}$.
Applying these in Eq.~\eqref{eq:WBT} is sufficient to remove
all phases, while preserving the overall texture, 
thus showing these models are also physically equivalent.
The calculations for the models 
$(\theta,\theta,0,\pi-2\theta)$,
$(\theta,\theta,\theta,\pi-2\theta)$,
$(\pi/4,\pi/4,0,\pi/2)$, and
$(\pi/4,\pi/4,\pi/4,\pi/2)$ are exactly the same with the
substitution $W_{dR}\leftrightarrow W_{uR}$.
With a similar analysis, one can also prove the equivalence
between the $(\theta,\theta,2\theta,2\theta)$, $(\theta,\theta,\pi-2\theta,\pi-2\theta)$, and
$(\pi/4,\pi/4,\pi/2,\pi/2)$ models, and between the 
$(\theta,2\theta,\theta,\theta)$, $(\theta,\pi-2\theta,\theta,\theta)$, and
$(\pi/4,\pi/2,\pi/4,\pi/4)$ models, but none of these were
found to be able to fit the data well.

In Table~\ref{tab:fitmodel} we present all the models found by~\cite{Bree:2024edl}
that are physically different and provided a good fit of the data.
This is our main result.
\begin{table}[htp!]
    \centering
    \begin{tabular}{c|c|c|c|c}
     \hline
\hline
\multirow{2}{*}{$(\theta,\alpha,\beta,\gamma)$}
     & \multirow{2}{*}{Range for $\theta$} 
     & \# of real param. & \# of real param. & Symmetry of\\
     &  & in down-type Yukawa & in up-type Yukawa & scalar potential\\
\hline
\\[-0.9em]
$(0,0,0,0)$ & 0 & 27 & 27 & CPa
\\[0.2em]
\hline
\\[-0.9em]
$(\frac{\pi}{2},0,0,\frac{\pi}{2})$ & \multirow{7}{*}{$\frac{\pi}{2}$} & $9^\diamond$ & 15  &
\\[0.2em]
$(\frac{\pi}{2},0,\frac{\pi}{2},0)$ &  & 15 & $9^\diamond$ &
\\[0.2em]
$(\frac{\pi}{2},0,\frac{\pi}{2},\frac{\pi}{2})$ &  & 15 & 15 &
\\[0.2em]
$(\frac{\pi}{2},\frac{\pi}{2},0,0)$ &  & 15 & 15 & CPb
\\[0.2em]
$(\frac{\pi}{2},\frac{\pi}{2},0,\frac{\pi}{2})$ &  & 15 & 13 &
\\[0.2em]
$(\frac{\pi}{2},\frac{\pi}{2},\frac{\pi}{2},0)$ &   & 13& 15 &
\\[0.2em]
$(\frac{\pi}{2},\frac{\pi}{2},\frac{\pi}{2},\frac{\pi}{2})$ &   & 13& 13 &
\\[0.2em]
\hline
\\[-0.9em]
$(\frac{\pi}{3},\frac{\pi}{3},0,\frac{\pi}{3})$ &  \multirow{3}{*}{$\frac{\pi}{3}$} & 9 & 9 &
\\[0.2em]
$(\frac{\pi}{3},\frac{\pi}{3},\frac{\pi}{3},0)$ &   & 9 & 9 & CPc
\\[0.2em]
$(\frac{\pi}{3},\frac{\pi}{3},\frac{\pi}{3},\frac{\pi}{3})$ &   & 9 & 9 &
\\[0.2em]

\hline
\\[-0.9em]
$(\theta,0,0,\theta)$ & \multirow{4}{*}{$\left(0, \frac{\pi}{2}\right)$}  & $9^\diamond$ & $9$ &
\\[0.2em]
$(\theta,0,\theta,0)$ &   & $9$& $9^\diamond$ & CPc ($\theta= \pi/3$)
\\[0.2em]
$(\theta,0,\theta,\theta)$ &   & $9$& $9$ & or
\\[0.2em]
$(\theta,\theta,0,0)$ &   & $9$& $9$ & CPd ($\theta \neq \pi/3$)
\\[0.2em]
\hline
\\[-0.9em]
$(\theta,\theta,0,\theta)$ & \multirow{3}{*}{$\left(0, \frac{\pi}{2}\right)\setminus\{\frac{\pi}{3}\}$}   & $9$ & $7$&  \\[0.2em]
$(\theta,\theta,\theta,0)$ &  & $7$ & $9$ & CPd
\\[0.2em]
$(\theta,\theta,\theta,\theta)$ &   & $7$& $7$ &
\\[0.2em]
\hline
\\[-0.9em]
$(\theta,\theta,0,2\theta)$ &  \multirow{4}{*}{ $\left(0, \frac{\pi}{2}\right)\setminus\{\frac{\pi}{3}\}$} & $9$ & $5$ &
\\[0.2em]
$(\theta,\theta,2\theta,0)$ &  & $5$ & $9$ & CPd
\\[0.2em]
$(\theta,\theta,\theta,2\theta)$ & & $7$ & $5$ &
\\[0.2em]
$(\theta,\theta,2\theta,\theta)$ &  & $5$ & $7$ &
\\[0.2em]
\hline
\hline
\end{tabular}
\caption{\label{tab:fitmodel}All 22 physically distinct
GCP-symmetric 3HDM compatible
with non zero and non-degenerate quark masses as well as a
non-vanishing Jarlskog invariant that provide a good fit of the data.}
\end{table}

\subsection{Parametrization}

In this analysis, we considered that the scalar fields can acquire 
generic complex vevs.
The motivation is that these symmetries are often used with soft symmetry
breaking terms in the scalar potential, thus allowing for the most general vevs.
As such,
\begin{equation}
\Phi_i
=
\left(
\begin{array}{c}
\phi^+_i\\
\frac{|v_i|e^{i\delta_i}}{\sqrt{2}} + \frac{1}{\sqrt{2}}(x_i + ix_{i+3})
\end{array}
\right)\, .
\end{equation}
We can use the gauge freedom to absorb one of the phases in the vevs,
for which we chose $\delta_1=0$. Therefore, we can parametrize
the vector of vevs as 
\begin{equation}
\vec{v}
=
(|v_1|,|v_2|e^{i\delta_2},|v_3|e^{i\delta_3})
\, .
\end{equation}
This vev contributes with four free parameters to our model. The last parameter
is constrained by the mass of the gauge bosons of the SM, such that
\begin{equation}
v^2 \equiv |v_1|^2 + |v_2|^2 + |v_3|^2 \simeq (246\text{ GeV})^2\, .
\end{equation}
The vev can also be parametrized as
\begin{equation}
\label{eq:polarparam}
\vec{v}
=
v(\cos({\beta_1})\sin({\beta_2}),\sin({\beta_1})\sin({\beta_2})e^{i\delta_2},
\cos({\beta_2})e^{i\delta_3})
\, .
\end{equation}

Looking at the CKM matrix, it was found by Branco and Lavoura \cite{Branco:1987mj}
that the magnitudes of the CKM matrix elements can be obtained by calculating
the traces of appropriate powers of the matrices $H_u$ and $H_d$. 
They observe that
\begin{equation}
\label{eq:trhhl}
\text{Tr}\left( H_u^a\, H_d^b\right)
\equiv
L_{ab}
=
\sum_{k, i}U_{ki}(D_u^{2})^a_{kk}(D_d^2)^b_{ii}
\, ,
\end{equation}
where $U_{ki}=|V_{ki}|^2$,
where $V$ is the CKM matrix. $V$ is unitary,
and therefore $U$ has only four independent entries. Thus, in order to
compute $U$, one only needs
\begin{eqnarray}
L_{11}&=U_{ki}(D^2_u)_{kk}(D_d^2)_{ii}\, ,
\nonumber\\
L_{12}&=U_{ki}(D^2_u)_{kk}(D_d^4)_{ii}\, ,
\nonumber\\
L_{21}&=U_{ki}(D_u^4)_{kk}(D_d^2)_{ii}\, ,
\nonumber\\
L_{22}&=U_{ki}(D_u^4)_{kk}(D_d^4)_{ii}\, .
\end{eqnarray}
These equations are linear in $U_{ki}$ and, therefore, invertible.
By picking $U_{11}$, $U_{13}$, $U_{21}$, and $U_{23}$ 
(equal to $|V_{11}|^2$, $|V_{13}|^2$, $|V_{21}|^2$, and $|V_{23}|^2$ 
respectively), 
we are able to find unique solutions for the magnitudes of the CKM elements
as functions of $L_{ab}$ and the quark masses \cite{Bree:2023ojl}
\begin{eqnarray}
\label{eq:Uij}
U_{11}&=&\left(m^2_b-m^2_s\right)\left(m^2_c-m^2_t\right)\, \frac{a_{11}}{\text{det}}\, ,
\nonumber\\
U_{13}&=&\left(m^2_d-m^2_s\right)\left(m^2_c-m^2_t\right)\, \frac{a_{13}}{\text{det}}\, ,
\nonumber\\
U_{21}&=&\left(m^2_b-m^2_s\right)\left(m^2_u-m^2_t\right)\, \frac{a_{21}}{\text{det}}\, ,
\nonumber\\
U_{23}&=&\left(m^2_d-m^2_s\right)\left(m^2_u-m^2_t\right)\, \frac{a_{23}}{\text{det}}\, ,
\end{eqnarray}
where
\begin{eqnarray}
a_{11}&=& L_{11}\left(m_b^2+m_s^2\right)\left(m_c^2+m_t^2\right) - L_{12}\left(m_c^2+m_t^2\right)
-L_{21}\left(m_b^2+m_s^2\right) + L_{22}
\nonumber\\
&-& m_b^2\left(m_c^2m_t^2\left(m_d^2+m_s^2\right) + m_s^2m_u^2
\left(m_c^2+m_t^2- m_u^2\right)\right) 
+ m_c^2m_d^2m_t^2\left(m_d^2-m_s^2\right) \, ,\\
a_{13}&=& -L_{11}\left( m_d^2+m_s^2 \right)\left( m_c^2+m_t^2 \right) + 
L_{12}\left( m_c^2+m_t^2 \right) + L_{21}\left( m_d^2+m_s^2 \right) -L_{22}
\nonumber\\
&+& m_b^2m_c^2m_t^2\left( m_d^2+m_s^2-m_b^2 \right) + m_d^2m_s^2\left( 
m_c^2\left( m_t^2+m_u^2 \right) +m_u^2\left( m_t^2-m_u^2 \right) \right)
\, ,\\
a_{21}&=& -L_{11}\left( m_b^2+m_s^2 \right)\left( m_u^2+m_t^2 \right) + 
L_{12}\left( m_u^2+m_t^2 \right) + L_{21}\left( m_b^2+m_s^2 \right) -L_{22}
\nonumber\\
&+& m_b^2\left(m_c^2m_s^2\left(m_u^2+m_t^2-m_c^2\right) + m_t^2m_u^2\left(m_d^2+m_s^2\right)\right) 
+ m_d^2m_t^2m_u^2\left(m_s^2-m_d^2\right)
\, ,\\
a_{23}&=& L_{11}\left( m_d^2+m_s^2 \right)\left( m_u^2+m_t^2 \right) - 
L_{12}\left( m_u^2+m_t^2 \right) - L_{21}\left( m_d^2+m_s^2 \right) +L_{22}
\nonumber\\
&+& m_t^2m_u^2\left( m_b^4-m_b^2\left( m_d^2+m_s^2 \right) -m_d^2m_s^2 \right)
+ m_c^2m_d^2m_s^2\left( m_c^2-m_u^2-m_t^2 \right)
\, ,
\end{eqnarray}
and
\begin{equation}
    \text{det} = \left( m_b^2-m_d^2 \right)\left( m_c^2-m_u^2 \right)
    \left( m_d^2-m_s^2 \right)\left( m_u^2-m_t^2 \right)
    \left( m_b^2-m_s^2 \right)\left( m_c^2-m_t^2 \right)\,.
\end{equation}
In these equations, the $L_{ij}$ are obtained by evaluating the left-hand side of Eq.~\eqref{eq:trhhl}. 

From unitarity we have that $|V_{32}|=|V_{11}V_{23} - V_{13}V_{21} |$.
Using the relation $|a-b|^2=|a|^2+|b|^2-2\text{Re}(ab^*)$, we find
\begin{equation}
\begin{split}
    R\equiv \text{Re}(V_{11}V_{23}V_{13}^*V_{21}^*) &= \frac{1}{2}\left(|V_{11}V_{23}|^2 + |V_{13}V_{21}|^2-|V_{11}V_{23} - V_{13}V_{21} |^2\right)\\
    &= \frac{1}{2}\left(U_{11}U_{23} + U_{13}U_{21} -U_{32}\right)\\
    &= \frac{1}{2}\left(1- U_{11} -U_{13}-U_{21}-U_{23}+ U_{11}U_{23} + U_{13}U_{21}\right)\, ,
\end{split}
\label{eq: R}
\end{equation}
where in the last equation we used 
$U_{12}+U_{22}+U_{32}=U_{11}+U_{12}+U_{13}=U_{21}+U_{22}+U_{23}=1$ 
to write $U_{32}$ as a function of $U_{11}$, $U_{13}$, $U_{21}$, and $U_{23}$.
Since
\begin{equation}
\label{eq:j2}
\begin{split}
    J_{CP}^2= \left[\text{Im}(V_{11}V_{23}V_{13}^*V_{21}^*)\right]^2 =
    U_{11}U_{23}U_{13}U_{23} - R^2 \, ,
\end{split}
\end{equation}
it is possible to determine the Jarlskog invariant $J_{CP}$ \cite{Jarlskog:1985ht}, up to its sign, with only these four CKM magnitudes.
Thus, up to some phase convention, we are also able to determine all
the CKM matrix elements' phases.

\subsection{Fitting Procedure}

We would like to fit 10 observables, 6 quark masses and 4 CKM elements,
with the parameters of our models, those being
$\beta_1$, $\beta_2$, $\delta_2$, and $\delta_3$, from our parametrization
in Eq.~\eqref{eq:polarparam}, as well as the parameters 
in the Yukawa textures for each one of our models \cite{Bree:2024edl}.

We performed a $\chi^2$ minimization of our model, using the 
CERN Minuit library \cite{James:1975dr}.
The observables relevant to this analysis, their mean experimental values,
and the experimental error, which, when both left and right
bounds are stated, is assumed to be the largest of the two,
are shown in Table~\ref{tab:fitobservables}.
The data on the quark masses, the CKM matrix elements, and the Jarlskog invariant 
experimental values were obtained from \cite{ParticleDataGroup:2024cfk}.
To obtain the magnitudes we used the fitted values for
$s_{12}= 0.22501\pm0.00068 $, $s_{13}=0.003732\pm0.000090 $,
$s_{23}= 0.04183\pm0.00079 $, and $\delta= 1.147\pm0.026$,
and plugged them into the unitary parametrization of the CKM matrix \cite{ParticleDataGroup:2024cfk}.

\begin{table}[h]
    \centering
    \begin{tabular}{c|c}
        Observable & Experimental Value \\
        \hline
\\[-0.9em]
        $m_u$ [MeV] & $2.16\pm0.07$ \\[0.1em]
        $m_c$ [MeV] & $1273.0\pm4.6$\\[0.1em]
        $m_t$ [GeV] & $172.57\pm0.29$\\[0.1em]
        $m_d$ [MeV] & $4.70\pm0.07$ \\[0.1em]
        $m_s$ [MeV] & $93.5\pm0.8$\\[0.1em]
        $m_b$ [MeV] & $4183\pm7$\\[0.1em]
        $|V_{11}|$ & $0.97434967\pm0.00016$ \\[0.1em]
        $|V_{13}|$ & $0.003732\pm0.000090$\\[0.1em]
        $|V_{21}|$ & $0.2248756\pm0.00068$ \\[0.1em]
        $|V_{23}|$ & $0.0418297\pm0.00079$ \\[0.1em]
        $J_{CP}$ & $(3.12\pm0.13)\times 10^{-5}$\\
        \hline
    \end{tabular}
    \caption{Experimental values. The quark masses were taken directly from \cite{ParticleDataGroup:2024cfk}. 
    The CKM magnitudes were obtained using the values 
    of $s_{12}$, $s_{13}$, $s_{23}$, and $\delta$, as well as the unitary parametrization of the CKM matrix \cite{ParticleDataGroup:2024cfk}.}
    \label{tab:fitobservables}
\end{table}

The $\chi^2$ function depends on the parameters of our model,
and is written as
\begin{equation}
\label{eq: chi2}
    \chi^2(\textbf{p}) = \sum_{i=1}^{10} \left(\frac{P_i(\textbf{p})-\overline{X_i}}{\sigma_i}\right)^2\, ,
\end{equation}
where $P_i(\textbf{p})$ is our model's prediction for the square of each of the
10 observables (excluding $J_{CP}$) in Table~\ref{tab:fitobservables},
$\overline{X_i}$ is the square of each experimental value,
and $\sigma_i$ is obtained via propagation of error. We start by calculating
$H_d$ and $H_u$, as well as their eigenvalues, which depend on 
$\beta_1$, $\beta_2$, $\delta_2$, $\delta_3$, and the Yukawa parameters.
Then, we evaluate the $L_{ij}$ with the left-hand side of Eq.~\eqref{eq:trhhl},
and the CKM elements are obtained with Eq.~\eqref{eq:Uij}. At the end, we compute
the value of $|J_{CP}|$ using the square root of the right-hand side of 
Eq.~\eqref{eq:j2} with  
the obtained values for $U_{11}$, $U_{13}$, $U_{21}$, and $U_{23}$.

Finally, it is important to note that the fits presented
are not the only possibilities. We found
several combinations of parameters that yield similar $\chi^2$ values, 
and chose for each model some specific example where it was smallest.

\subsection{CPc Models}
\label{section fit}

Here we list all unique Yukawa textures for all possible
CPc invariant models, as well as a numerical fit for all parameters
in the model.
Onwards, every parameter $\{x_{ij}\}$ is a real parameter.

\subsubsection{$(\theta, \alpha, \beta, \gamma) = ( \pi/3, \pi/3, 0,\pi/3)$}
\ba
&&
\Gamma_1 =
\left(
\begin{array}{ccc}
 a_{11} & a_{12} & a_{13} \\
 a_{21} & a_{22} & a_{23} \\
 0 & 0 & 0
\end{array}
\right),
\ \ \
\Gamma_2 =
\left(
\begin{array}{ccc}
 -a_{21} & -a_{22} & -a_{23} \\
 a_{11} & a_{12} & a_{13} \\
 0 & 0 & 0
\end{array}
\right),
\ \ \
\Gamma_3 =
\left(
\begin{array}{ccc}
 0 & 0 & 0 \\
 0 & 0 & 0 \\
 c_{31} & c_{32} & c_{33}
\end{array}
\right),
\nonumber \\[1em]
&&
\Delta_1 =
\left(
\begin{array}{ccc}
 i b_{11} & i b_{12} & b_{13} \\
 i b_{12} & -i b_{11} & b_{23} \\
 b_{31} & b_{32} & 0
\end{array}
\right),
\ \ \
\Delta_2 =
\left(
\begin{array}{ccc}
 i b_{12} & -i b_{11} & -b_{23} \\
 -i b_{11} & -i b_{12} & b_{13} \\
 -b_{32} & b_{31} & 0
\end{array}
\right),
\ \ \
\Delta_3 =
\left(
\begin{array}{ccc}
d_{11} & d_{12} & 0 \\
-d_{12} & d_{11} & 0 \\
0 & 0 & d_{33}
\end{array}
\right).
\nonumber \\
\ea

\begin{table}[h!]
    \centering
\begin{tabular}{|c|c|c|}
\hline
{$a_{11}=    8.9412\times10^{-4}$} & 
{$a_{12}=   -4.4566\times10^{-4}$} & 
{$a_{13}=   -5.1798\times10^{-4}$} \\
\hline
{$a_{21}=    4.7028\times10^{-4}$} & 
{$a_{22}=   -3.2118\times10^{-4}$} & 
{$a_{23}=   -3.5121\times10^{-4}$} \\
\hline
{$b_{11}=   -2.4902\times10^{-3}$} & 
{$b_{12}=    5.5300\times10^{-3}$} & 
{$b_{13}=   -8.6619\times10^{-4}$} \\
\hline
{$b_{23}=   -3.6586\times10^{-4}$} & 
{$b_{31}=   -0.2696$} & 
{$b_{32}=   -0.9653$} \\
\hline
{$c_{31}=   -6.9372\times10^{-2}$} & 
{$c_{32}=    8.1668\times10^{-2}$ } & 
{$c_{33}=    0.1129$} \\
\hline
{$d_{11}=    2.7292\times10^{-2}$} & 
{$d_{12}=    2.7090\times10^{-3}$ } & 
{$d_{33}=   -0.38833$} \\
\hline
\end{tabular}
\\
\begin{tabular}{|c|c|}
\hline
{$\beta_1=    0.35499$} &
{$\beta_2=    1.41587$} \\
\hline
{$\delta_2=   6.1747$} &
{$\delta_3=   4.9430$} \\
\hline
\end{tabular}
    \caption{Numerical fit for the $(\pi/3,\pi/3,0,\pi/3)$ model.}
    \label{tab:fit1}
\end{table}


\begin{table}[h!]
    \centering
\begin{tabular}{| c | c | c || c | c |}
\hline
{$v_1=       227.899\gev$} & 
{$v_2=       84.482\gev$} & 
{$v_3=       37.961\gev$} &
{$|V_{11}|=      0.9743496$} &
{$|V_{13}|=      0.00373198 $} \\
\hline
{$m_u=     2.160\mev$} & 
{$m_c=     1273.000\mev$} & 
{$m_t=   172.569954\gev$} &
{$|V_{21}|=      0.2248755$} &
{$|V_{23}|=      0.0418299$} \\
\hline
{$m_d=     4.700\mev$} & 
{$m_s=     93.500\mev$} & 
{$m_b=     4183.001\mev$} &
{$J_{CP}=        3.1205\times10^{-5}$} &
{$\chi^2=     4.242\times10^{-7}$} \\
\hline
\end{tabular}
    \caption{Predictions for the $(\pi/3,\pi/3,0,\pi/3)$ model.}
    \label{tab:results1}
\end{table}

\subsubsection{$(\theta, \alpha, \beta, \gamma) = ( \pi/3, \pi/3, \pi/3, 0)$}
\ba
&&
\Gamma_1 =
\left(
\begin{array}{ccc}
 i a_{11} & i a_{12} & a_{13} \\
 i a_{12} & -i a_{11} & a_{23} \\
 a_{31} & a_{32} & 0
\end{array}
\right),
\ \ \
\Gamma_2 =
\left(
\begin{array}{ccc}
 i a_{12} & -i a_{11} & -a_{23} \\
 -i a_{11} & -i a_{12} & a_{13} \\
 -a_{32} & a_{31} & 0
\end{array}
\right),
\ \ \
\Gamma_3 =
\left(
\begin{array}{ccc}
c_{11} & c_{12} & 0 \\
-c_{12} & c_{11} & 0 \\
0 & 0 & c_{33}
\end{array}
\right),
\nonumber \\[1em]
&&
\Delta_1 =
\left(
\begin{array}{ccc}
 b_{11} & b_{12} & b_{13} \\
 b_{21} & b_{22} & b_{23} \\
 0 & 0 & 0
\end{array}
\right),
\ \ \
\Delta_2 =
\left(
\begin{array}{ccc}
 -b_{21} & -b_{22} & -b_{23} \\
 b_{11} & b_{12} & b_{13} \\
 0 & 0 & 0
\end{array}
\right),
\ \ \
\Delta_3 =
\left(
\begin{array}{ccc}
 0 & 0 & 0 \\
 0 & 0 & 0 \\
 d_{31} & d_{32} & d_{33}
\end{array}
\right).
\ea

\begin {table}[h!]
    \centering
\begin {tabular} {| c | c | c |}
\hline
{$a_{11}=   -1.6646\times10^{-3}$} &
{$a_{12}=    2.3703\times10^{-3}$} &
{$a_{13}=   -1.3647\times10^{-3}$} \\
\hline
{$a_{23}=    4.7637\times10^{-2}$} &
{$a_{31}=    1.2680\times10^{-4}$} &
{$a_{32}=    5.2205\times10^{-4}$} \\
\hline
{$b_{11}=    1.8691\times10^{-2}$} &
{$b_{12}=    8.9354\times10^{-3}$} &
{$b_{13}=    6.7407\times10^{-4}$} \\
\hline
{$b_{21}=    1.735206$} &
{$b_{22}=   -0.192215$} &
{$b_{23}=   -0.928780$} \\
\hline
{$c_{11}=    1.2818\times10^{-3}$} &
{$c_{12}=    3.5882\times10^{-4}$} &
{$c_{33}=   -1.6011\times10^{-4}$} \\
\hline
{$d_{31}=   -4.53495\times10^{-3}$} &
{$d_{32}=   -1.2470\times10^{-3}$} &
{$d_{33}=    7.5176\times10^{-4}$} \\
\hline
\end {tabular}
\\
\begin {tabular} {| c | c |}
\hline
{$\beta_1=    1.5647$} &
{$\beta_2=    0.52552$} \\
\hline
{$\delta_2=   4.4654$} &
{$\delta_3=   5.9611$} \\
\hline
\end {tabular}
    \caption {Numerical fit for the $ (\pi/3, \pi/3, \pi/3, 0) $ model.}
    \label {tab : fit2}
\end {table}


\begin {table}[h!]
    \centering
\begin {tabular} {| c | c | c || c | c |}
\hline
{$v_1=       0.746\gev$} &
{$v_2=       123.407\gev$} &
{$v_3=       212.806\gev$} &
{$|V_{11}|=      0.9743498$} &
{$|V_{13}|=      0.00373201$} \\
\hline
{$m_u=     2.160\mev$} &
{$m_c=     1273.000\mev$} &
{$m_t=   172.569958\gev$} &
{$|V_{21}|=      0.2248762$} &
{$|V_{23}|=      0.0418298$} \\
\hline
{$m_d=     4.700\mev$} &
{$m_s=     93.500\mev$} &
{$m_b=     4183.002\mev$} &
{$J_{CP}=        3.1040\times10^{-5}$} &
{$\chi^2=     1.831\times10^{-6}$} \\
\hline
\end {tabular}
    \caption {Predictions for the $ (\pi/3, \pi/3, \pi/3, 0) $ model.}
    \label {tab : results2}
\end {table}

\subsubsection{$(\theta, \alpha, \beta,\gamma) = ( \pi/3, \pi/3, \pi/3, \pi/3)$}
\ba
&&
\Gamma_1 =
\left(
\begin{array}{ccc}
 i a_{11} & i a_{12} & a_{13} \\
 i a_{12} & -i a_{11} & a_{23} \\
 a_{31} & a_{32} & 0
\end{array}
\right),
\ \ \
\Gamma_2 =
\left(
\begin{array}{ccc}
 i a_{12} & -i a_{11} & -a_{23} \\
 -i a_{11} & -i a_{12} & a_{13} \\
 -a_{32} & a_{31} & 0
\end{array}
\right),
\ \ \
\Gamma_3 =
\left(
\begin{array}{ccc}
c_{11} & c_{12} & 0 \\
-c_{12} & c_{11} & 0 \\
0 & 0 & c_{33}
\end{array}
\right),
\nonumber \\[1em]
&&
\Delta_1 =
\left(
\begin{array}{ccc}
 i b_{11} & i b_{12} & b_{13} \\
 i b_{12} & -i b_{11} & b_{23} \\
 b_{31} & b_{32} & 0
\end{array}
\right),
\ \ \
\Delta_2 =
\left(
\begin{array}{ccc}
 i b_{12} & -i b_{11} & -b_{23} \\
 -i b_{11} & -i b_{12} & b_{13} \\
 -b_{32} & b_{31} & 0
\end{array}
\right),
\ \ \
\Delta_3 =
\left(
\begin{array}{ccc}
d_{11} & d_{12} & 0 \\
-d_{12} & d_{11} & 0 \\
0 & 0 & d_{33}
\end{array}
\right).
\nonumber \\
\ea

\begin {table}[h!]
    \centering
\begin {tabular} {| c | c | c |}
\hline
{$a_{11}=   -3.8671\times10^{-3}$} &
{$a_{12}=   -1.5107\times10^{-3}$} &
{$a_{13}=    4.1153\times10^{-5}$} \\
\hline
{$a_{23}=    3.1061\times10^{-4}$} &
{$a_{31}=    1.3404\times10^{-2}$} &
{$a_{32}=   -1.8965\times10^{-2}$} \\
\hline
{$b_{11}=    2.4636\times10^{-3}$} &
{$b_{12}=   -1.1650\times10^{-4}$} &
{$b_{13}=   -1.3690\times10^{-2}$} \\
\hline
{$b_{23}=   -0.3869$} &
{$b_{31}=    2.1427\times10^{-3}$} &
{$b_{32}=    3.1146\times10^{-2}$} \\
\hline
{$c_{11}=   -1.5044\times10^{-2}$} &
{$c_{12}=   -9.9010\times10^{-3}$} &
{$c_{33}=   -3.3660\times10^{-3}$} \\
\hline
{$d_{11}=   -9.9090\times10^{-3}$} &
{$d_{12}=    2.3680\times10^{-3}$} &
{$d_{33}=   -3.94896$} \\
\hline
\end {tabular}
\\
\begin {tabular} {| c | c |}
\hline
{$\beta_1=    6.5791\times10^{-2}$} &
{$\beta_2=    1.33635$} \\
\hline
{$\delta_2=   3.58289$} &
{$\delta_3=   4.68078$} \\
\hline
\end {tabular}
    \caption {Numerical fit for the $ (\pi/3, \pi/3, \pi/3, \pi/3) $ model.}
    \label {tab : fit3}
\end {table}


\begin {table}[h!]
    \centering
\begin {tabular} {| c | c | c || c | c |}
\hline
{$v_1=       238.753\gev$} &
{$v_2=       15.731\gev$} &
{$v_3=       57.147\gev$} &
{$|V_{11}|=      0.9743498$} &
{$|V_{13}|=      0.003732$} \\
\hline
{$m_u=     2.160\mev$} &
{$m_c=     1273.002\mev$} &
{$m_t=   172.570031\gev$} &
{$|V_{21}|=      0.224876$} &
{$|V_{23}|=      0.0418299$} \\
\hline
{$m_d=     4.700\mev$} &
{$m_s=     93.500\mev$} &
{$m_b=     4182.999\mev$} &
{$J_{CP}=        3.1084\times10^{-5}$} &
{$\chi^2=     1.121\times10^{-6}$} \\
\hline
\end {tabular}
    \caption {Predictions for the $ (\pi/3, \pi/3, \pi/3, \pi/3) $ model.}
    \label {tab : results3}
\end {table}

The next four models are invariant simultaneously under CPc ($\theta= \pi/3$)
and CPd ($\theta \neq \pi/3$),
since the allowed range for $\theta$ is $(0,\pi/2)$.

\subsubsection{$\theta \in \left(0,\pi/2\right)\ \ \&\ \  
(\alpha, \beta, \gamma) = (0, 0,\theta)$}
\ba
&&
\Gamma_1 = \Gamma_2 = 0,
\ \ \
\Gamma_3 = \left(
\begin{array}{ccc}
 c_{11} & c_{12} & c_{13} \\
 c_{21} & c_{22} & c_{23} \\
 c_{31} & c_{32} & c_{33}
\end{array}
\right),
\nonumber \\[1em]
&&
\Delta_1 = 
\left(
\begin{array}{ccc}
 b_{11} & b_{12} & 0 \\
 b_{21} & b_{22} & 0 \\
 b_{31} & b_{32} & 0
\end{array}
\right),
\ \ \
\Delta_2 =
\left(
\begin{array}{ccc}
 -b_{12} & b_{11} & 0 \\
 -b_{22} & b_{21} & 0 \\
 -b_{32} & b_{31} & 0
\end{array}
\right),
\ \ \
\Delta_3 =
\left(
\begin{array}{ccc}
 0 & 0 & d_{13} \\
 0 & 0 & d_{23} \\
 0 & 0 & d_{33}
\end{array}
\right).
\ea

\begin {table}[h!]
    \centering
\begin {tabular} {| c | c | c |}
\hline
{$b_{11}=    1.8132\times10^{-2}$} &
{$b_{12}=   -5.8503\times10^{-3}$} &
{$b_{21}=    0.22389$} \\
\hline
{$b_{22}=    0.93627$} &
{$b_{31}=    8.8927\times10^{-2}$} &
{$b_{32}=    0.223827$} \\
\hline
{$c_{11}=    2.8468\times10^{-3}$} &
{$c_{12}=    2.92355\times10^{-3}$} &
{$c_{13}=   -4.3198\times10^{-3}$} \\
\hline
{$c_{21}=   -0.21657$} &
{$c_{22}=   -0.21050$} &
{$c_{23}=   -0.73252$} \\
\hline
{$c_{31}=   -4.8236\times10^{-2}$} &
{$c_{32}=   -4.1874\times10^{-2}$} &
{$c_{33}=   -2.0134$} \\
\hline
{$d_{13}=   -1.6376\times10^{-2}$} &
{$d_{23}=    3.8566\times10^{-2}$} &
{$d_{33}=   -2.0484\times10^{-2}$} \\
\hline
\end {tabular}
\\
\begin {tabular} {| c | c |}
\hline
{$\beta_1=    0.86576$} &
{$\beta_2=    1.54146$} \\
\hline
{$\delta_2=   4.6118$} &
{$\delta_3=   5.6518$} \\
\hline
\end {tabular}
    \caption {Numerical fit for the $ (\pi/3, 0, 0, \pi/3) $ model.}
    \label {tab : fit4}
\end {table}


\begin {table}[h!]
    \centering
\begin {tabular} {| c | c | c || c | c |}
\hline
{$v_1=       159.354\gev$} &
{$v_2=       187.270\gev$} &
{$v_3=       7.216\gev$} &
{$|V_{11}|=      0.9743496$} &
{$|V_{13}|=      0.00373195$} \\
\hline
{$m_u=     2.160\mev$} &
{$m_c=     1273.001\mev$} &
{$m_t=   172.570020\gev$} &
{$|V_{21}|=      0.2248752$} &
{$|V_{23}|=     0.0418299$} \\
\hline
{$m_d=     4.700\mev$} &
{$m_s=     93.500\mev$} &
{$m_b=     4182.999\mev$} &
{$J_{CP} =        3.1240\times10^{-5}$} &
{$\chi^2=     1.082\times10^{-6}$} \\
\hline
\end {tabular}
    \caption {Predictions for the $ (\pi/3, 0, 0, \pi/3) $ model.}
    \label {tab : results4}
\end {table}

\subsubsection{$\theta \in \left(0,\pi/2\right)\ \ \&\ \  
(\alpha, \beta, \gamma) = (0,\theta, 0)$}
\ba
&&
\Gamma_1 = 
\left(
\begin{array}{ccc}
 a_{11} & a_{12} & 0 \\
 a_{21} & a_{22} & 0 \\
 a_{31} & a_{32} & 0
\end{array}
\right),
\ \ \
\Gamma_2 =
\left(
\begin{array}{ccc}
 -a_{12} & a_{11} & 0 \\
 -a_{22} & a_{21} & 0 \\
 -a_{32} & a_{31} & 0
\end{array}
\right),
\ \ \
\Gamma_3 =
\left(
\begin{array}{ccc}
 0 & 0 & c_{13} \\
 0 & 0 & c_{23} \\
 0 & 0 & c_{33}
\end{array}
\right),
\nonumber \\[1em]
&&
\Delta_1 = \Delta_2 = 0,
\ \ \
\Delta_3 = \left(
\begin{array}{ccc}
 d_{11} & d_{12} & d_{13} \\
 d_{21} & d_{22} & d_{23} \\
 d_{31} & d_{32} & d_{33}
\end{array}
\right).
\ea

\begin {table}[h!]
    \centering
\begin {tabular} {| c | c | c |}
\hline
{$a_{11}=   -1.9618\times10^{-2}$} &
{$a_{12}=    8.0505\times10^{-3}$} &
{$a_{21}=   -5.5453\times10^{-3}$} \\
\hline
{$a_{22}=    2.8723\times10^{-3}$} &
{$a_{31}=   -1.1916\times10^{-3}$} &
{$a_{32}=    8.1412\times10^{-4}$} \\
\hline
{$c_{13}=    10.8424$} &
{$c_{23}=    2.8890$} &
{$c_{33}=    0.608560$} \\
\hline
{$d_{11}=   -1.2768\times10^{2}$} &
{$d_{12}=   -1.1359\times10^{3}$} &
{$d_{13}=   -1.4923\times10^{2}$} \\
\hline
{$d_{21}=   -26.6641$} &
{$d_{22}=   -2.8378\times10^{2}$} &
{$d_{23}=   -30.5568$} \\
\hline
{$d_{31}=   -5.5901$} &
{$d_{32}=   -62.0842$} &
{$d_{33}=   -6.3773$} \\
\hline
\end {tabular}
\\
\begin {tabular} {| c | c |}
\hline
{$\beta_1=    0.71843$} &
{$\beta_2=    1.56996$} \\
\hline
{$\delta_2=   5.57146$} &
{$\delta_3=   0.33487$} \\
\hline
\end {tabular}
    \caption {Numerical fit for the $ (\pi/3, 0, \pi/3, 0) $ model.}
    \label {tab : fit5}
\end {table}


\begin {table}[h!]
    \centering
\begin {tabular} {| c | c | c || c | c |}
\hline
{$v_1=       185.198\gev$} &
{$v_2=       161.919\gev$} &
{$v_3=       0.205\gev$} &
{$|V_{11}|=      0.9743498$} &
{$|V_{13}|=      0.00373202$} \\
\hline
{$m_u=     2.160\mev$} &
{$m_c=     1272.999\mev$} &
{$m_t=   172.570036\gev$} &
{$|V_{21}|=      0.2248762$} &
{$|V_{23}|=      0.0418293$} \\
\hline
{$m_d=     4.700\mev$} &
{$m_s=     93.501\mev$} &
{$m_b=     4183.001\mev$} &
{$J_{CP} =        3.1058\times10^{-5}$} &
{$\chi^2=     2.109\times10^{-6}$} \\
\hline
\end {tabular}
    \caption {Predictions for the $ (\pi/3, 0, \pi/3, 0) $ model.}
    \label {tab : results5}
\end {table}

\subsubsection{$\theta \in \left(0,\pi/2\right)\ \ \&\ \  
(\alpha, \beta, \gamma) = (0,\theta,\theta)$}
\ba
&&
\Gamma_1 = 
\left(
\begin{array}{ccc}
 a_{11} & a_{12} & 0 \\
 a_{21} & a_{22} & 0 \\
 a_{31} & a_{32} & 0
\end{array}
\right),
\ \ \
\Gamma_2 =
\left(
\begin{array}{ccc}
 -a_{12} & a_{11} & 0 \\
 -a_{22} & a_{21} & 0 \\
 -a_{32} & a_{31} & 0
\end{array}
\right),
\ \ \
\Gamma_3 =
\left(
\begin{array}{ccc}
 0 & 0 & c_{13} \\
 0 & 0 & c_{23} \\
 0 & 0 & c_{33}
\end{array}
\right),
\nonumber \\[1em]
&&
\Delta_1 = 
\left(
\begin{array}{ccc}
 b_{11} & b_{12} & 0 \\
 b_{21} & b_{22} & 0 \\
 b_{31} & b_{32} & 0
\end{array}
\right),
\ \ \
\Delta_2 =
\left(
\begin{array}{ccc}
 -b_{12} & b_{11} & 0 \\
 -b_{22} & b_{21} & 0 \\
 -b_{32} & b_{31} & 0
\end{array}
\right),
\ \ \
\Delta_3 =
\left(
\begin{array}{ccc}
 0 & 0 & d_{13} \\
 0 & 0 & d_{23} \\
 0 & 0 & d_{33}
\end{array}
\right).
\ea

\begin {table}[h!]
    \centering
\begin {tabular} {| c | c | c |}
\hline
{$a_{11}=   -1.0250\times10^{-3}$} &
{$a_{12}=   -5.3896\times10^{-4}$} &
{$a_{21}=   -6.2337\times10^{-4}$} \\
\hline
{$a_{22}=    1.0402\times10^{-3}$} &
{$a_{31}=    4.6064\times10^{-3}$} &
{$a_{32}=    2.1503\times10^{-2}$} \\
\hline
{$b_{11}=    2.1795\times10^{-2}$} &
{$b_{12}=    3.9809\times10^{-3}$} &
{$b_{21}=   -8.3727\times10^{-3}$} \\
\hline
{$b_{22}=    3.0239\times10^{-2}$} &
{$b_{31}=    0.11180$} &
{$b_{32}=    0.990376$} \\
\hline
{$c_{13}=   -7.3580\times10^{-4}$} &
{$c_{23}=    2.1469\times10^{-3}$} &
{$c_{33}=    9.4615\times10^{-2}$} \\
\hline
{$d_{13}=   -5.8608\times10^{-2}$} &
{$d_{23}=    2.8031\times10^{-2}$} &
{$d_{33}=   -0.14465$} \\
\hline
\end {tabular}
\\
\begin {tabular} {| c | c |}
\hline
{$\beta_1=    0.74080$} &
{$\beta_2=    1.46630$} \\
\hline
{$\delta_2=   4.6641$} &
{$\delta_3=   1.2498$} \\
\hline
\end {tabular}
    \caption {Numerical fit for the $ (\pi/3, 0, \pi/3, \pi/3) $ model.}
    \label {tab : fit6}
\end {table}


\begin {table}[h!]
    \centering
\begin {tabular} {| c | c | c || c | c |}
\hline
{$v_1=       180.541\gev$} &
{$v_2=       165.114\gev$} &
{$v_3=       25.660\gev$} &
{$|V_{11}|=      0.9743495$} &
{$|V_{13}|=      0.00373196$} \\
\hline
{$m_u=     2.160\mev$} &
{$m_c=     1272.999\mev$} &
{$m_t=   172.569887\gev$} &
{$|V_{21}|=      0.2248751$} &
{$|V_{23}|=      0.0418306$} \\
\hline
{$m_d=     4.700\mev$} &
{$m_s=     93.501\mev$} &
{$m_b=     4183.001\mev$} &
{$J_{CP} =        3.1306\times10^{-5}$} &
{$\chi^2=     4.689\times10^{-6}$} \\
\hline
\end {tabular}
    \caption {Predictions for the $ (\pi/3, 0, \pi/3, \pi/3) $ model.}
    \label {tab : results6}
\end {table}

\subsubsection{$\theta \in \left(0,\pi/2\right)\ \ \&\ \  
(\alpha, \beta, \gamma) = (\theta,0,0)$}
\ba
&&
\Gamma_1 =
\left(
\begin{array}{ccc}
 a_{11} & a_{12} & a_{13} \\
 a_{21} & a_{22} & a_{23} \\
 0 & 0 & 0
\end{array}
\right),
\ \ \
\Gamma_2 =
\left(
\begin{array}{ccc}
 -a_{21} & -a_{22} & -a_{23} \\
 a_{11} & a_{12} & a_{13} \\
 0 & 0 & 0
\end{array}
\right),
\ \ \
\Gamma_3 =
\left(
\begin{array}{ccc}
 0 & 0 & 0 \\
 0 & 0 & 0 \\
 c_{31} & c_{32} & c_{33}
\end{array}
\right),
\nonumber \\[1em]
&&
\Delta_1 =
\left(
\begin{array}{ccc}
 b_{11} & b_{12} & b_{13} \\
 b_{21} & b_{22} & b_{23} \\
 0 & 0 & 0
\end{array}
\right),
\ \ \
\Delta_2 =
\left(
\begin{array}{ccc}
 -b_{21} & -b_{22} & -b_{23} \\
 b_{11} & b_{12} & b_{13} \\
 0 & 0 & 0
\end{array}
\right),
\ \ \
\Delta_3 =
\left(
\begin{array}{ccc}
 0 & 0 & 0 \\
 0 & 0 & 0 \\
 d_{31} & d_{32} & d_{33}
\end{array}
\right). \nonumber \\
\ea

\begin {table}[h!]
    \centering
\begin {tabular} {| c | c | c |}
\hline
{$a_{11}=    7.7672\times10^{-2}$} &
{$a_{12}=    1.2058\times10^{-2}$} &
{$a_{13}=    5.5174\times10^{-2}$} \\
\hline
{$a_{21}=    6.5458\times10^{-2}$} &
{$a_{22}=    7.0297\times10^{-3}$} &
{$a_{23}=    4.8166\times10^{-2}$} \\
\hline
{$b_{11}=    1.7522$} &
{$b_{12}=    0.14067$} &
{$b_{13}=    3.6597$} \\
\hline
{$b_{21}=    1.3590$} &
{$b_{22}=    0.12970$} &
{$b_{23}=    2.9441$} \\
\hline
{$c_{31}=    2.2441\times10^{-4}$} &
{$c_{32}=   -1.1225\times10^{-4}$} &
{$c_{33}=    2.7791\times10^{-4}$} \\
\hline
{$d_{31}=   -2.2147\times10^{-3}$} &
{$d_{32}=   -5.1854\times10^{-4}$} &
{$d_{33}=    6.3062\times10^{-3}$} \\
\hline
\end {tabular}
\\
\begin {tabular} {| c | c |}
\hline
{$\beta_1=    1.55374$} &
{$\beta_2=    0.19205$} \\
\hline
{$\delta_2=   1.00391$} &
{$\delta_3=   2.66119$} \\
\hline
\end {tabular}
    \caption {Numerical fit for the $ (\pi/3, \pi/3, 0, 0) $ model.}
    \label {tab : fit7}
\end {table}


\begin {table}[h!]
    \centering
\begin {tabular} {| c | c | c || c | c |}
\hline
{$v_1=       0.801\gev$} &
{$v_2=       46.948\gev$} &
{$v_3=       241.477\gev$} &
{$|V_{11}|=      0.9743497$} &
{$|V_{13}|=      0.003732$} \\
\hline
{$m_u=     2.160\mev$} &
{$m_c=     1272.999\mev$} &
{$m_t=   172.570034\gev$} &
{$|V_{21}|=      0.2248755$} &
{$|V_{23}|=      0.0418296$} \\
\hline
{$m_d=     4.700\mev$} &
{$m_s=     93.499\mev$} &
{$m_b=     4183.001\mev$} &
{$J_{CP} =        3.1178\times10^{-5}$} &
{$\chi^2=     9.034\times10^{-7}$} \\
\hline
\end {tabular}
    \caption {Predictions for the $ (\pi/3, \pi/3, 0, 0) $ model.}
    \label {tab : results7}
\end {table}

\subsection{CPd Models with 14 parameters}
There are a total of 8 models 
(although only 6 are physically distinct), all of them CPd invariant,
that have only 14 parameters,
the lowest amount out of all 40 models.
They are the models in Table~\ref{tab:fixedmodel} and 
Table~\ref{tab:rangemodel} that contain 5 parameters in both the 
up and down-type Yukawa matrices,
i.e. the $(\pi/5,2\pi/5,\pi/5,2\pi/5)$,
$(\pi/5,2\pi/5,2\pi/5,\pi/5)$,
$(2\pi/5,\pi/5,2\pi/5,\pi/5)$,
$(2\pi/5,\pi/5,\pi/5,2\pi/5)$,
$(\theta,\theta,2\theta,2\theta)$,
$(\theta,2\theta,\theta,\theta)$,
$(\theta,\theta,\pi-2\theta,\pi-2\theta)$, and
$(\theta,\pi-2\theta,\theta,\theta)$ models 
(the last two being physically the same as the previous two).

However, we could not find a good fit that correctly reproduced the
experimental data. We can conclude that these models are too stringent,
and thus not viable.

\subsection{CPd Models with 16 parameters}
Here we list the Yukawa textures for the two
physically distinct
CPd invariant models that contain only 16 parameters,
as well as a numerical fit for all parameters in the model.

\subsubsection{$\theta \in \left(0,\pi/2\right)\setminus\{\pi/3\}\ \ \&\ \  
(\alpha, \beta, \gamma) = (\theta, \theta,2\theta)$}
\ba
&&
\Gamma_1 =
\left(
\begin{array}{ccc}
 0 & 0 & a_{13} \\
 0 & 0 & a_{23} \\
 a_{31} & a_{32} & 0
\end{array}
\right),
\ \ \
\Gamma_2 =
\left(
\begin{array}{ccc}
 0 & 0 & -a_{23} \\
 0 & 0 & a_{13} \\
 -a_{32} & a_{31} & 0
\end{array}
\right),
\ \ \
\Gamma_3 =
\left(
\begin{array}{ccc}
 c_{11} & c_{12} & 0 \\
 -c_{12} & c_{11} & 0 \\
 0 & 0 & c_{33}
\end{array}
\right),
\nonumber \\[1em]
&&
\Delta_1 =
\left(
\begin{array}{ccc}
 b_{11} & b_{12} & b_{13} \\
 -b_{12} & b_{11} & b_{23} \\
 0 & 0 & 0
\end{array}
\right),
\ \ \
\Delta_2 =
\left(
\begin{array}{ccc}
 -b_{12} & b_{11} & -b_{23} \\
 -b_{11} & -b_{12} & b_{13} \\
 0 & 0 & 0
\end{array}
\right),
\ \ \
\Delta_3 =
\left(
\begin{array}{ccc}
 0 & 0 & 0 \\
 0 & 0 & 0 \\
 0 & 0 & d_{33}
\end{array}
\right). \nonumber \\
\ea

\begin {table}[h!]
    \centering
\begin {tabular} {| c | c | c |}
\hline
{$a_{13}=   -2.3664\times10^{-2}$} &
{$a_{23}=   -4.2447\times10^{-3}$} &
{$a_{31}=   -5.6320\times10^{-5}$} \\
\hline
{$a_{32}=   -1.1021\times10^{-4}$} &
{$b_{11}=    5.9528\times10^{-3}$} &
{$b_{12}=   -4.2570\times10^{-3}$} \\
\hline
{$b_{13}=    0.968288$} &
{$b_{23}=    0.21582$} &
{$c_{11}=    0.27008$} \\
\hline
{$c_{12}=   -0.32282$} &
{$c_{33}=    6.6957\times10^{-2}$} &
{$d_{33}=   -1.3577$} \\
\hline
\end {tabular}
\\
\begin {tabular} {| c | c |}
\hline
{$\beta_1=    1.57000$} &
{$\beta_2=    1.56955$} \\
\hline
{$\delta_2=   5.23284$} &
{$\delta_3=   1.34920$} \\
\hline
\end {tabular}
    \caption {Numerical fit for the $ (\theta, \theta, \theta, 2\theta) $ model, with $\theta\in(0,\pi/2)\setminus\{\pi/3\}$.}
    \label {tab : fit8}
\end {table}


\begin {table}[h!]
    \centering
\begin {tabular} {| c | c | c || c | c |}
\hline
{$v_1=       0.196\gev$} &
{$v_2=       245.9997\gev$} &
{$v_3=       0.306\gev$} &
{$|V_{11}|=      0.9743497$} &
{$|V_{13}|=      0.003732$} \\
\hline
{$m_u=     2.160\mev$} &
{$m_c=     1273.001\mev$} &
{$m_t=   172.570046\gev$} &
{$|V_{21}|=      0.2248755$} &
{$|V_{23}|=      0.0418299$} \\
\hline
{$m_d=     4.700\mev$} &
{$m_s=     93.500\mev$} &
{$m_b=     4183.003\mev$} &
{$J_{CP}=        3.1172\times10^{-5}$} &
{$\chi^2=     7.882\times10^{-7}$} \\
\hline
\end {tabular}
    \caption {Predictions for the $ (\theta, \theta, \theta, 2\theta) $ model, with $\theta\in(0,\pi/2)\setminus\{\pi/3\}$.}
    \label {tab : results8.}
\end {table}

\newpage
\subsubsection{$\theta \in \left(0,\pi/2\right)\setminus\{\pi/3\}\ \ \&\ \  (\alpha, \beta, \gamma) = (\theta,2\theta, \theta)$}
\ba
&&
\Gamma_1 =
\left(
\begin{array}{ccc}
 a_{11} & a_{12} & a_{13} \\
 -a_{12} & a_{11} & a_{23} \\
 0 & 0 & 0
\end{array}
\right),
\ \ \
\Gamma_2 =
\left(
\begin{array}{ccc}
 -a_{12} & a_{11} & -a_{23} \\
 -a_{11} & -a_{12} & a_{13} \\
 0 & 0 & 0
\end{array}
\right),
\ \ \
\Gamma_3 =
\left(
\begin{array}{ccc}
 0 & 0 & 0 \\
 0 & 0 & 0 \\
 0 & 0 & c_{33}
\end{array}
\right)
,
\nonumber \\[1em]
&&
\Delta_1 =
\left(
\begin{array}{ccc}
 0 & 0 & b_{13} \\
 0 & 0 & b_{23} \\
 b_{31} & b_{32} & 0
\end{array}
\right),
\ \ \
\Delta_2 =
\left(
\begin{array}{ccc}
 0 & 0 & -b_{23} \\
 0 & 0 & b_{13} \\
 -b_{32} & b_{31} & 0
\end{array}
\right),
\ \ \
\Delta_3 =
\left(
\begin{array}{ccc}
 d_{11} & d_{12} & 0 \\
 -d_{12} & d_{11} & 0 \\
 0 & 0 & d_{33}
\end{array}
\right). \nonumber \\
\ea

\begin {table}[h!]
    \centering
\begin {tabular} {| c | c | c |}
\hline
{$a_{11}=   -4.9234\times10^{-4}$} &
{$a_{12}=   -2.1762\times10^{-4}$} &
{$a_{13}=   -1.8535\times10^{-2}$} \\
\hline
{$a_{23}=   -1.5326\times10^{-2}$} &
{$b_{13}=   -0.78743$} &
{$b_{23}=   -0.60467$} \\
\hline
{$b_{31}=    1.4584\times10^{-3}$} &
{$b_{32}=    8.5660\times10^{-4}$} &
{$c_{33}=   -2.1144\times10^{-2}$} \\
\hline
{$d_{11}=    7.5057\times10^{-3}$} &
{$d_{12}=   -0.12438$} &
{$d_{33}=   -0.72325$} \\
\hline
\end {tabular}
\\
\begin {tabular} {| c | c |}
\hline
{$\beta_1=    1.5217$} &
{$\beta_2=    1.5136$} \\
\hline
{$\delta_2=   3.3458$} &
{$\delta_3=   3.3363$} \\
\hline
\end {tabular}
    \caption {Numerical fit for the $ (\theta, \theta, 2\theta, \theta) $ model, with $\theta\in(0,\pi/2)\setminus\{\pi/3\}$.}
    \label {tab : fit9}
\end {table}


\begin {table}[h!]
    \centering
\begin {tabular} {| c | c | c || c | c |}
\hline
{$v_1=       12.044\gev$} &
{$v_2=       245.302\gev$} &
{$v_3=       14.072\gev$} &
{$|V_{11}|=      0.9743498$} &
{$|V_{13}|=      0.00373194$} \\
\hline
{$m_u=     2.160\mev$} &
{$m_c=     1272.996\mev$} &
{$m_t=   172.570368\gev$} &
{$|V_{21}|=      0.2248754$} &
{$|V_{23}|=      0.0418292$} \\
\hline
{$m_d=     4.700\mev$} &
{$m_s=     93.499\mev$} &
{$m_b=     4183.005\mev$} &
{$J_{CP}=        3.1124\times10^{-5}$} &
{$\chi^2=     5.686\times10^{-6}$} \\
\hline
\end {tabular}
    \caption {Predictions for the $ (\theta, \theta, 2\theta, \theta) $ model, with $\theta\in(0,\pi/2)\setminus\{\pi/3\}$.}
    \label {tab : results9}
\end {table}

\subsection{CPd Models with 18 parameters}
We can split the models with 18 parameters into two groups.
The first contains 5 parameters in one of the Yukawa matrix, and 9
parameters in the other.
The second contains 7 parameters in each.

We verified that all ``5+9'' models are able to fit the data very well.
Next we present a numerical fit for one of these models.

\subsubsection{$\theta \in \left(0,\pi/2\right)\setminus\{\pi/3\}\ \ \&\ \  (\alpha, \beta, \gamma) = (\theta,2\theta, 0)$}
\ba
&&
\Gamma_1 =
\left(
\begin{array}{ccc}
 a_{11} &a_{12} & a_{13} \\
 -a_{12} &a_{11} & a_{23} \\
 0 & 0 & 0
\end{array}
\right),
\ \ \
\Gamma_2 =
\left(
\begin{array}{ccc}
 -a_{12} &a_{11} & -a_{23} \\
 -a_{11} &-a_{12} & a_{13} \\
 0 & 0 & 0
\end{array}
\right),
\ \ \
\Gamma_3 =
\left(
\begin{array}{ccc}
 0 & 0 & 0 \\
 0 & 0 & 0 \\
 0 & 0 & c_{33}
\end{array}
\right)
,
\nonumber \\[1em]
&&
\Delta_1 =
\left(
\begin{array}{ccc}
 b_{11} & b_{12} & b_{13} \\
 b_{21} & b_{22} & b_{23} \\
 0 & 0 & 0
\end{array}
\right),
\ \ \
\Delta_2 =
\left(
\begin{array}{ccc}
 -b_{21} & -b_{22} & -b_{23} \\
 b_{11} & b_{12} & b_{13} \\
 0 & 0 & 0
\end{array}
\right),
\ \ \
\Delta_3 =
\left(
\begin{array}{ccc}
 0 & 0 & 0 \\
 0 & 0 & 0 \\
 d_{31} & d_{32} & d_{33}
\end{array}
\right). \nonumber \\
\ea

\begin {table}[h!]
    \centering
\begin{tabular}{|c|c|c|c|}
\hline
{$a_{11}=    4.2317\times10^{-4}$} &
{$a_{12}=   -3.3175\times10^{-4}$} &
{$a_{13}=    2.3687\times10^{-2}$} &
{$a_{23}=    3.9871\times10^{-3}$} \\
\hline
\end{tabular}
\\
\begin {tabular} {| c | c | c |}
\hline
{$b_{11}=    0.284428$} &
{$b_{12}=    0.931148$} &
{$b_{13}=    0.140346$} \\
\hline
{$b_{21}=    3.0755\times10^{-2}$} &
{$b_{22}=    0.120658$} &
{$b_{23}=    1.3688\times10^{-2}$} \\
\hline
\end {tabular}
\\
\begin{tabular}{|c|c|c|c|}
\hline
{$c_{33}=    4.2981\times10^{-2}$} &
{$d_{31}=    0.381379$} &
{$d_{32}=    1.4155$} &
{$d_{33}=    0.175183$} \\
\hline
\end{tabular}
\\
\begin {tabular} {| c | c |}
\hline
{$\beta_1=    1.5663$} &
{$\beta_2=    1.5427$} \\
\hline
{$\delta_2=   1.9131$} &
{$\delta_3=   1.8953$} \\
\hline
\end {tabular}
    \caption {Numerical fit for the $ (\theta, \theta,2\theta, 0) $ model, with $\theta\in(0,\pi/2)\setminus\{\pi/3\}$.}
    \label {tab : fit12}
\end {table}


\begin {table}[h!]
    \centering
\begin {tabular} {| c | c | c || c | c |}
\hline
{$v_1=       1.095\gev$} &
{$v_2=       245.900\gev$} &
{$v_3=       6.919\gev$} &
{$|V_{11}|=      0.9743498$} &
{$|V_{13}|=      0.00373204$} \\
\hline
{$m_u=     2.160\mev$} &
{$m_c=     1272.999\mev$} &
{$m_t=   172.570136\gev$} &
{$|V_{21}|=      0.2248763$} &
{$|V_{23}|=      0.0418290$} \\
\hline
{$m_d=     4.700\mev$} &
{$m_s=     93.500\mev$} &
{$m_b=     4182.997\mev$} &
{$J_{CP}=        3.1043\times10^{-5}$} &
{$\chi^2=     3.960\times10^{-6}$} \\
\hline
\end {tabular}
    \caption {Predictions for the $ (\theta, \theta,2\theta, 0) $ model, with $\theta\in(0,\pi/2)\setminus\{\pi/3\}$.}
    \label {tab : results12}
\end {table}

Next we present a numerical fit for the $(\theta,\theta,\theta,\theta)$ model,
with $\theta\in\left(0,\pi/2\right)\setminus\{\pi/3\}$.
Note that the fact that this model provides a good fit is a sufficient condition
for it being also true for the $(\pi/3,\pi/3,\pi/3,\pi/3)$ model
(which we had already verified),
since this is a boundary model of the previous one.

\subsubsection{$\theta \in \left(0,\pi/2\right)\setminus\{\pi/3\}\ \ \&\ \  (\alpha, \beta, \gamma) = (\theta,\theta, \theta)$}
\ba
&&
\Gamma_1 =
\left(
\begin{array}{ccc}
 0 & 0 & a_{13} \\
 0 & 0 & a_{23} \\
 a_{31} & a_{32} & 0
\end{array}
\right),
\ \ \
\Gamma_2 =
\left(
\begin{array}{ccc}
 0 & 0 & -a_{23} \\
 0 & 0 & a_{13} \\
 -a_{32} & a_{31} & 0
\end{array}
\right),
\ \ \
\Gamma_3 =
\left(
\begin{array}{ccc}
c_{11} & c_{12} & 0 \\
-c_{12} & c_{11} & 0 \\
0 & 0 & c_{33}
\end{array}
\right),
\nonumber \\[1em]
&&
\Delta_1 =
\left(
\begin{array}{ccc}
 0 & 0 & b_{13} \\
 0 & 0 & b_{23} \\
 b_{31} & b_{32} & 0
\end{array}
\right),
\ \ \
\Delta_2 =
\left(
\begin{array}{ccc}
 0 & 0 & -b_{23} \\
 0 & 0 & b_{13} \\
 -b_{32} & b_{31} & 0
\end{array}
\right),
\ \ \
\Delta_3 =
\left(
\begin{array}{ccc}
d_{11} & d_{12} & 0 \\
-d_{12} & d_{11} & 0 \\
0 & 0 & d_{33}
\end{array}
\right).
\nonumber \\
\ea

\begin {table}[h!]
    \centering
\begin{tabular}{|c|c|c|c|}
\hline
{$a_{13}=    1.1352\times10^{-3}$} &
{$a_{23}=    4.6385\times10^{-5}$} &
{$a_{31}=    8.3951\times10^{-3}$} &
{$a_{32}=    7.0686\times10^{-3}$} \\
\hline
{$b_{13}=   -7.2731\times10^{-3}$} &
{$b_{23}=   -8.3814\times10^{-4}$} &
{$b_{31}=    0.12179$} &
{$b_{32}=    0.98391$} \\
\hline
\end{tabular}
\\
\begin {tabular} {| c | c | c |}
\hline
{$c_{11}=   -4.5029\times10^{-3}$} &
{$c_{12}=    3.9284\times10^{-3}$} &
{$c_{33}=    1.26866$} \\
\hline
{$d_{11}=    1.4319\times10^{-2}$} &
{$d_{12}=   -2.7088\times10^{-3}$} &
{$d_{33}=   -2.36874$} \\
\hline
\end {tabular}
\\
\begin {tabular} {| c | c |}
\hline
{$\beta_1=    5.7763\times10^{-2}$} &
{$\beta_2=    1.5539$} \\
\hline
{$\delta_2=   2.5391$} &
{$\delta_3=   3.8464$} \\
\hline
\end {tabular}
    \caption {Numerical fit for the $ (\theta, \theta,\theta, \theta) $ model, with $\theta\in(0,\pi/2)\setminus\{\pi/3\}$.}
    \label {tab : fit13}
\end {table}


\begin {table}[h!]
    \centering
\begin {tabular} {| c | c | c || c | c |}
\hline
{$v_1=       245.555\gev$} &
{$v_2=       14.200\gev$} &
{$v_3=       4.145\gev$} &
{$|V_{11}|=      0.9743495$} &
{$|V_{13}|=      0.00373197$} \\
\hline
{$m_u=     2.160\mev$} &
{$m_c=     1272.999\mev$} &
{$m_t=   172.570049\gev$} &
{$|V_{21}|=      0.2248745$} &
{$|V_{23}|=      0.0418305$} \\
\hline
{$m_d=     4.700\mev$} &
{$m_s=     93.501\mev$} &
{$m_b=     4183.002\mev$} &
{$J_{CP}=        3.1340\times10^{-5}$} &
{$\chi^2=     6.594\times10^{-6}$} \\
\hline
\end {tabular}
    \caption {Predictions for the $ (\theta, \theta,\theta, \theta) $ model, with $\theta\in(0,\pi/2)\setminus\{\pi/3\}$.}
    \label {tab : results13}
\end {table}

The remaining 18 parameter models are all models where every angle
has a fixed value, namely $\theta=\pi/4$.
As we have proved, each of these is physically equivalent to
another model, but with fewer parameters.
We have already discussed the fits with equivalent models with fewer parameters.
Given the equivalence,
we can easily identify which models do (or not) provide a good fit.
Despite that, we have verified explicitly that, even in the basis where the
Yukawa matrices have 18 parameters,
neither the 
$(\pi/4,\pi/4,\pi/2,\pi/2)$ nor the
$(\pi/4,\pi/2,\pi/4,\pi/4)$ were able
to provide a good fit.

\subsection{CPd Models with 20 parameters}

There are 4 models that contain only 20 parameters. Out of these,
the $(\pi/4,\pi/4,0,\pi/2)$ and $(\pi/4,\pi/4,\pi/2,0)$ models
are physically equivalent to the
$(\theta,\theta,0,2\theta)$ and 
$(\theta,\theta,2\theta,0)$
models, respectively,
as well as being boundary models.
Thus, these are all able to fit the data.
The remaining ones, $(\theta,\theta,0,\theta)$ and 
$(\theta,\theta,\theta,0)$, are also able to provide a good fit. 
Next we present a numerical fit for the latter.

\subsubsection{$\theta \in \left(0,\pi/2\right)\setminus\{\pi/3\}\ \ \&\ \  (\alpha, \beta, \gamma) = (\theta,\theta, 0)$}
\ba
&&
\Gamma_1 =
\left(
\begin{array}{ccc}
 0 & 0 & a_{13} \\
 0 & 0 & a_{23} \\
 a_{31} & a_{32} & 0
\end{array}
\right),
\ \ \
\Gamma_2 =
\left(
\begin{array}{ccc}
 0 & 0 & -a_{23} \\
 0 & 0 & a_{13} \\
 -a_{32} & a_{31} & 0
\end{array}
\right),
\ \ \
\Gamma_3 =
\left(
\begin{array}{ccc}
c_{11} & c_{12} & 0 \\
-c_{12} & c_{11} & 0 \\
0 & 0 & c_{33}
\end{array}
\right),
\nonumber \\[1em]
&&
\Delta_1 =
\left(
\begin{array}{ccc}
 b_{11} & b_{12} & b_{13} \\
 b_{21} & b_{22} & b_{23} \\
 0 & 0 & 0
\end{array}
\right),
\ \ \
\Delta_2 =
\left(
\begin{array}{ccc}
 -b_{21} & -b_{22} & -b_{23} \\
 b_{11} & b_{12} & b_{13} \\
 0 & 0 & 0
\end{array}
\right),
\ \ \
\Delta_3 =
\left(
\begin{array}{ccc}
 0 & 0 & 0 \\
 0 & 0 & 0 \\
 d_{31} & d_{32} & d_{33}
\end{array}
\right). \nonumber \\
\ea

\begin {table}[h!]
    \centering
\begin{tabular}{|c|c|c|c|}
\hline
{$a_{13}=    4.7445\times10^{-2}$} &
{$a_{23}=    3.5194\times10^{-3}$} &
{$a_{31}=    7.5597\times10^{-5}$} &
{$a_{32}=    2.8639\times10^{-4}$} \\
\hline
\end{tabular}
\\
\begin {tabular} {| c | c | c |}
\hline
{$b_{11}=    1.28840$} &
{$b_{12}=    1.22512$} &
{$b_{13}=    0.83014$} \\
\hline
{$b_{21}=    4.4989\times10^{-2}$} &
{$b_{22}=    4.6203\times10^{-2}$} &
{$b_{23}=    1.4354\times10^{-2}$} \\
\hline
{$c_{11}=   -5.2192\times10^{-4}$} &
{$c_{12}=    2.9395\times10^{-4}$} &
{$c_{33}=   -1.2519\times10^{-4}$} \\
\hline
{$d_{31}=    6.3500\times10^{-4}$} &
{$d_{32}=    5.2064\times10^{-4}$} &
{$d_{33}=    8.5741\times10^{-4}$} \\
\hline
\end {tabular}
\\
\begin {tabular} {| c | c |}
\hline
{$\beta_1=    0.625429$} &
{$\beta_2=    0.529768$} \\
\hline
{$\delta_2=   6.2763$} &
{$\delta_3=   2.6041$} \\
\hline
\end {tabular}
    \caption {Numerical fit for the $ (\theta, \theta,\theta, 0) $ model, with $\theta\in(0,\pi/2)\setminus\{\pi/3\}$.}
    \label {tab : fit14}
\end {table}


\begin {table}[h!]
    \centering
\begin {tabular} {| c | c | c || c | c |}
\hline
{$v_1=       100.781\gev$} &
{$v_2=       72.778\gev$} &
{$v_3=       212.279\gev$} &
{$|V_{11}|=      0.9743497$} &
{$|V_{13}|=      0.00373198$} \\
\hline
{$m_u=     2.160\mev$} &
{$m_c=     1273.000\mev$} &
{$m_t=   172.569947\gev$} &
{$|V_{21}|=      0.2248756$} &
{$|V_{23}|=      0.0418297$} \\
\hline
{$m_d=     4.700\mev$} &
{$m_s=     93.500\mev$} &
{$m_b=     4182.999\mev$} &
{$J_{CP}=        3.1169\times10^{-5}$} &
{$\chi^2=     2.215\times10^{-7}$} \\
\hline
\end {tabular}
    \caption {Predictions for the $ (\theta, \theta,\theta, 0) $
model, with $\theta\in(0,\pi/2)\setminus\{\pi/3\}$.}
    \label {tab : results14}
\end {table}

\subsection{CPb Models}
Every one of the models invariant under CPb, i.e. $\theta=\pi/2$, 
can be thought of as a boundary model to each of the first seven models
presented in Table~\ref{tab:rangemodel}, respectively.
Since we found viable fits for each of these, the corresponding boundary
also has a good fit. Note that, for example, the $(\pi/2,\pi/2,0,0)$
is simultaneous a boundary model for the $(\theta,\theta,0,0)$ and
$(\theta,\theta,\pi-2\theta,\pi-2\theta)$ models, but only the former
provided a good fit to the data.

\section{\label{sec:concl}Conclusions}
We studied 3HDM models with a softly-broken GCP symmetry,
attempting a fit to the 6 quark masses and the 4 quark mixing angles.
We started by showing that the 40 models of \cite{Bree:2024edl} are, in fact,
not all independent. We introduced the notion of boundary models
and showed the explicit basis transformations needed in order to relate equivalent models
(including models which apparently had different numbers of parameters).

We showed that, unlike in the 2HDM case, there are 3HDM GCP symmetric models
able to fit the experimental constraints from quark masses and mixings.
These fits are not unique, as there seems to be residual degrees of freedom
in parameter space that reproduce equally good predictions.
We found 22 models compatible with the observables,
listed in Table~\ref{tab:fitmodel}.
For each class of viable models,
we present one point in parameter space leading to a very good fit.

We verified that all CPc models, the 3HDM equivalent to the only
non trivial GCP symmetric 2HDM, fit very well the data, unlike 
what happened in the 2HDM. However, these models possess a large number
of free parameters, containing a total of 22 free parameters
(18 Yukawa parameters and 4 from the parametrization of the vevs).
In addition, we concluded that the trivial CPa model,
all possible CPb models and some CPd models can also fit the data.
In particular, CPd models with
16 parameters (12 in the Yukawa matrices and 4 from the vevs)
can fit the data.

In conclusion, the 3HDM with a GCP transformation is a viable model,
although it permits the existence of flavor changing neutral couplings,
which are constrained by experimental observations.
That study must be performed in detail model by model,
and lies beyond the scope of the present article.

\begin{acknowledgments}
\noindent
This work is supported in part by the Portuguese
Fundação para a Ciência e Tecnologia (FCT) through the PRR (Recovery and Resilience
Plan), within the scope of the investment "RE-C06-i06 - Science Plus
Capacity Building", measure "RE-C06-i06.m02 - Reinforcement of
financing for International Partnerships in Science,
Technology and Innovation of the PRR", under the project with
reference 2024.01362.CERN;
the work is
also supported by FCT under Contracts UIDB/00777/2020, and UIDP/00777/2020.
The FCT projects are partially funded through
POCTI (FEDER), COMPETE, QREN, and the EU.
\end{acknowledgments}

\vspace{5ex}


\bibliographystyle{JHEP}
\bibliography{ref}

@article{Ferreira:2010bm,
    author = "Ferreira, P. M. and Silva, Joao P.",
    title = "{A Two-Higgs Doublet Model With Remarkable CP Properties}",
    eprint = "1001.0574",
    archivePrefix = "arXiv",
    primaryClass = "hep-ph",
    doi = "10.1140/epjc/s10052-010-1384-5",
    journal = "Eur. Phys. J. C",
    volume = "69",
    pages = "45--52",
    year = "2010"
}

@article{Bree:2024edl,
    author = "Bree, Iris and Correia, Duarte D. and Silva, Joao P.",
    title = "{Generalized CP symmetries in three-Higgs-doublet models}",
    eprint = "2407.09615",
    archivePrefix = "arXiv",
    primaryClass = "hep-ph",
    doi = "10.1103/PhysRevD.110.035028",
    journal = "Phys. Rev. D",
    volume = "110",
    number = "3",
    pages = "035028",
    year = "2024"
}

@article{Botella:1994cs,
    author = "Botella, F. J. and Silva, Joao P.",
    title = "{Jarlskog - like invariants for theories with scalars and fermions}",
    eprint = "hep-ph/9411288",
    archivePrefix = "arXiv",
    reportNumber = "FTUV-94-68, IFIC-94-65",
    doi = "10.1103/PhysRevD.51.3870",
    journal = "Phys. Rev. D",
    volume = "51",
    pages = "3870--3875",
    year = "1995"
}

@book{Branco:1999fs,
    author = "Branco, Gustavo C. and Lavoura, Luis and Silva, Joao P.",
    title = "{CP Violation}",
    doi = "10.1093/oso/9780198503996.001.0001",
    isbn = "978-1-383-02075-5, 978-0-19-850399-6",
    volume = "103",
    year = "1999"
}

@article{Davidson:2005cw,
    author = "Davidson, Sacha and Haber, Howard E.",
    title = "{Basis-independent methods for the two-Higgs-doublet model}",
    eprint = "hep-ph/0504050",
    archivePrefix = "arXiv",
    reportNumber = "IPPP-03-23, DCPT-03-46, SCIPP-04-15",
    doi = "10.1103/PhysRevD.72.099902",
    journal = "Phys. Rev. D",
    volume = "72",
    pages = "035004",
    year = "2005",
    note = "[Erratum: Phys.Rev.D 72, 099902 (2005)]"
}

@article{Cabibbo:1963yz,
    author = "Cabibbo, Nicola",
    title = "{Unitary Symmetry and Leptonic Decays}",
    doi = "10.1103/PhysRevLett.10.531",
    journal = "Phys. Rev. Lett.",
    volume = "10",
    pages = "531--533",
    year = "1963"
}

@article{Kobayashi:1973fv,
    author = "Kobayashi, Makoto and Maskawa, Toshihide",
    title = "{CP Violation in the Renormalizable Theory of Weak Interaction}",
    reportNumber = "KUNS-242",
    doi = "10.1143/PTP.49.652",
    journal = "Prog. Theor. Phys.",
    volume = "49",
    pages = "652--657",
    year = "1973"
}

@article{Jarlskog:1985ht,
    author = "Jarlskog, C.",
    title = "{Commutator of the Quark Mass Matrices in the Standard Electroweak Model and a Measure of Maximal CP Nonconservation}",
    reportNumber = "USIP-85-14",
    doi = "10.1103/PhysRevLett.55.1039",
    journal = "Phys. Rev. Lett.",
    volume = "55",
    pages = "1039",
    year = "1985"
}

@article{Jarlskog:1985cw,
    author = "Jarlskog, C.",
    title = "{A Basis Independent Formulation of the Connection Between Quark Mass Matrices, CP Violation and Experiment}",
    reportNumber = "CERN-TH-4242/85",
    doi = "10.1007/BF01565198",
    journal = "Z. Phys. C",
    volume = "29",
    pages = "491--497",
    year = "1985"
}

@article{ParticleDataGroup:2024cfk,
    author = "Navas, S. and others",
    collaboration = "Particle Data Group",
    title = "{Review of particle physics}",
    doi = "10.1103/PhysRevD.110.030001",
    journal = "Phys. Rev. D",
    volume = "110",
    number = "3",
    pages = "030001",
    year = "2024"
}

@article{Bernabeu:1986fc,
    author = "Bernabeu, J. and Branco, G. C. and Gronau, M.",
    title = "{CP Restrictions on Quark Mass Matrices}",
    reportNumber = "CERN-TH-4329",
    doi = "10.1016/0370-2693(86)90659-3",
    journal = "Phys. Lett. B",
    volume = "169",
    pages = "243--247",
    year = "1986"
}

@article{Ecker:1987qp,
    author = "Ecker, G. and Grimus, W. and Neufeld, H.",
    title = "{A Standard Form for Generalized {CP} Transformations}",
    reportNumber = "UWThPh-1987-23",
    doi = "10.1088/0305-4470/20/12/010",
    journal = "J. Phys. A",
    volume = "20",
    pages = "L807",
    year = "1987"
}

@article{Branco:1987mj,
    author = "Branco, G. C. and Lavoura, L.",
    title = "{Rephasing Invariant Parametrization of the Quark Mixing Matrix}",
    reportNumber = "CERN-TH-4874/87",
    doi = "10.1016/0370-2693(88)91216-6",
    journal = "Phys. Lett. B",
    volume = "208",
    pages = "123--130",
    year = "1988"
}

@article{Bree:2023ojl,
    author = "Bree, Iris and Carrolo, Sergio and Romao, Jorge C. and Silva, Joao P.",
    title = "{A viable $A_4$ 3HDM theory of quark mass matrices}",
    eprint = "2301.04676",
    archivePrefix = "arXiv",
    primaryClass = "hep-ph",
    reportNumber = "CFTP/23-001",
    doi = "10.1140/epjc/s10052-023-11463-5",
    journal = "Eur. Phys. J. C",
    volume = "83",
    number = "4",
    pages = "292",
    year = "2023"
}

@article{James:1975dr,
    author = "James, F. and Roos, M.",
    title = "{Minuit: A System for Function Minimization and Analysis of the Parameter Errors and Correlations}",
    reportNumber = "CERN-DD-75-20",
    doi = "10.1016/0010-4655(75)90039-9",
    journal = "Comput. Phys. Commun.",
    volume = "10",
    pages = "343--367",
    year = "1975"
}

\pagebreak
\appendix
\section{\label{app:models}
Diagrammatic representation of the models 
}

\begin{figure}
    \centering
    \includegraphics[width=1\linewidth]{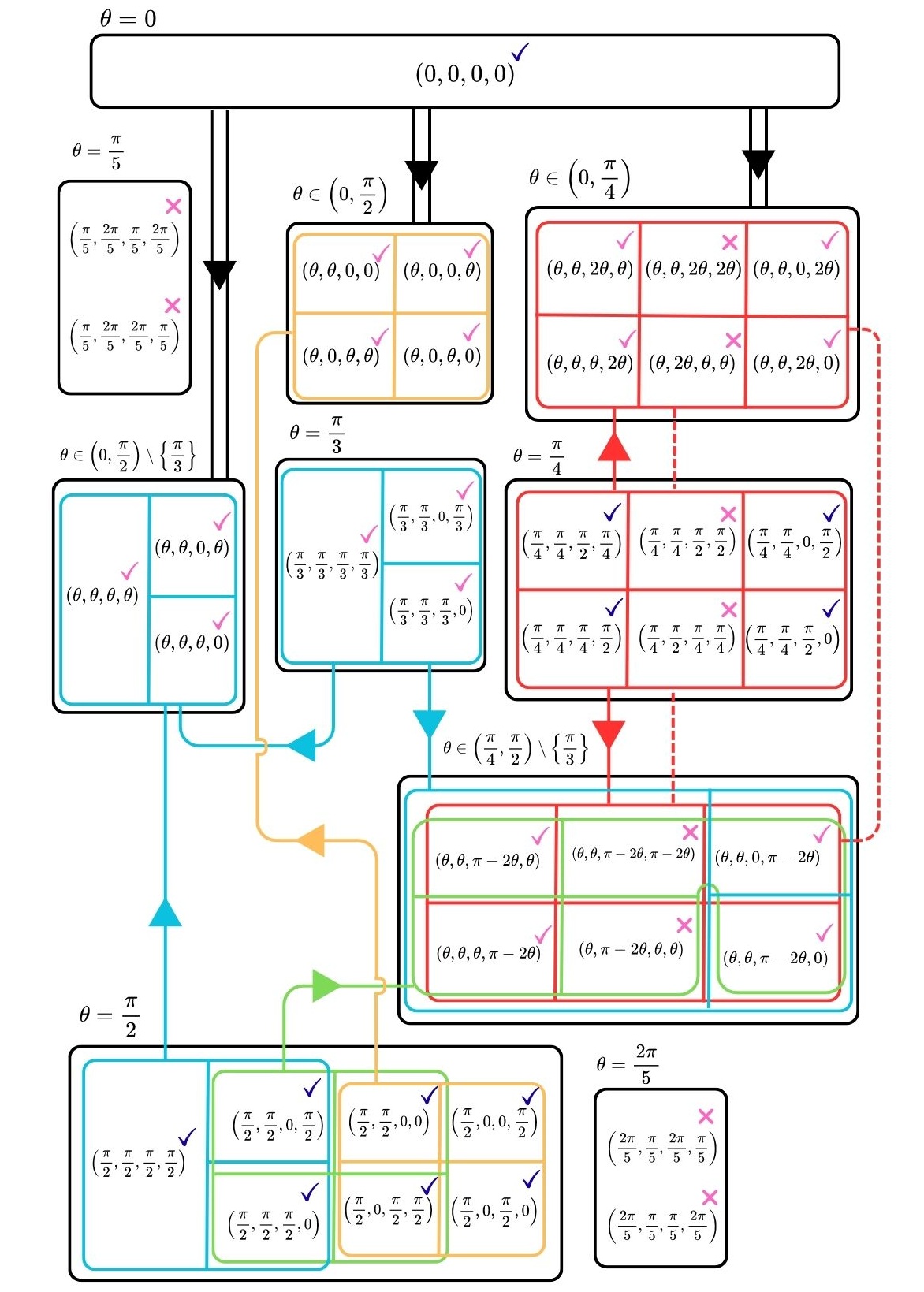}
    \caption{Diagrammatic representation of the models. 
    }
    \label{fig:models}
\end{figure}
Each
    of the models in Tables~\ref{tab:fixedmodel} and~\ref{tab:rangemodel} is represented in Fig.~\ref{fig:models}, in the notation
    $(\theta,\alpha,\beta,\gamma)$.
    A black box contains models where the allowed range/value for $\theta$ is the same, and the respective range/value
    appears above each box.
    A {\it boundary model}, i.e. a model whose parameter space
    is a superset of another model's parameter space, has an arrow pointing towards each of its subsets.
    A double lined arrow between black boxes means that the boundary
    model is a superset of every model inside the pointed at
    black box.
    A coloured arrow between coloured boxes relates the models
    only in the same relative position within each
    coloured box.
    A pink check (cross) mark next to a model means that
    a successful (failed) attempt to fit the experimental data
    was performed on that model.
    A blue check mark means that model is sure to be able
    to fit the data, since it is a superset/boundary model
    of at least
    one model that was found to fit the data.
    A coloured dashed line between coloured boxes means
    that the models in the same relative position are physically equivalent, i.e. the Yukawa textures are related by a unitary change of quark basis.

\end{document}